\documentclass[preprint,aps,nofootinbib]{revtex4-1}

\usepackage{color}
\usepackage{graphicx}
\usepackage{amsmath,slashed}
\usepackage{amsfonts}
\usepackage{amssymb}
\usepackage{hyperref}
\usepackage[capitalise]{cleveref}
\usepackage{epstopdf}
\usepackage{multirow}

\def\be{\begin{equation}}
\def\ee{\end{equation}}
\def\vep{\varepsilon}

\def\tr{\mathrm{tr}}
\def\ordr{\mathcal{O}}

\def\sib{\begin{align}r{\sigma}}
\def\<{\langle}
\def\>{\rangle}

\def\Res{{\text{Res}}}
\def\cycl{{\text{cycl}}}
\def\mt{{\mathcal{T}}}

\def\lag{{\mathcal{L}}}

\def\non{\nonumber}
\def\bea{\begin{eqnarray}}
\def\eea{\end{eqnarray}}
\def\bean{\begin{eqnarray*}}
\def\eean{\end{eqnarray*}}

\def\hatp{{\hat{p}}}
\def\hatm{{\hat{M}}}

\def\Adj{{\text{ad}}}
\def\wzw{{\text{wzw}}}
\def\SO{{\text{SO}}}
\def\SU{{\text{SU}}}
\def\U{{\text{U}}}

\def\hatP{{\hat{P}}}
\def\ctt{{\text{c}}}
\def\hats{{\hat{s}}}

\newcommand{\Rmnum}[1]{\expandafter\@slowromancap\romannumeral #1@}

\begin{document}

\vspace*{.3cm}

\title{Soft Bootstrap and Effective Field Theories  }

\author{\vspace{0.3cm} Ian Low$^{\, a,b}$ and Zhewei Yin$^{\, b}$}
\affiliation{\vspace{0.5cm}
\mbox{$^a$ High Energy Physics Division, Argonne National Laboratory, Lemont, IL 60439, USA}\\
\mbox{$^b$ Department of Physics and Astronomy, Northwestern University, Evanston, IL 60208, USA} 
 \vspace{0.5cm}
}

\begin{abstract}
The soft bootstrap program aims to construct consistent effective field theories (EFT's) by  recursively imposing the desired soft limit on tree-level scattering amplitudes through on-shell recursion relations.  A prime example is the  leading  two-derivative operator in the EFT of $\SU (N)\times \SU (N)/\SU (N)$ nonlinear sigma model (NLSM), where $\ordr (p^2)$ amplitudes with an arbitrary multiplicity of external particles can be soft-bootstrapped. We extend the program to $\ordr (p^4)$ operators and introduce the ``soft blocks,'' which are the seeds for soft bootstrap. The number of soft blocks coincides with the number of independent operators at a given order in the derivative expansion and the incalculable Wilson coefficient  emerges naturally.   We also uncover a new soft-constructible EFT involving the ``multi-trace'' operator  at the leading two-derivative order, which is matched to $\SO (N+1)/ \SO (N)$ NLSM. In addition, we consider Wess-Zumino-Witten (WZW) terms, the existence of which, or the lack thereof, depends on the number of flavors in the EFT, after a novel application of Bose symmetry.  Remarkably, we find agreements with  group-theoretic considerations on the existence of WZW terms in $\SU (N)$ NLSM for $N\ge 3$ and the absence of  WZW terms in $\SO (N)$ NLSM for $N\neq 5$.

\end{abstract}

\maketitle

\tableofcontents

\section{Introduction}

Soft bootstrap is a program  to construct consistent EFT's by recursively imposing the desired soft limit on tree-level amplitudes. The first attempt dates back almost half-a-century ago in the context of pions in low-energy QCD, whose scattering amplitudes exhibit vanishing soft behavior known as the Adler's zero condition \cite{Adler:1964um}. It was first shown by Susskind and Frye in Ref.~\cite{Susskind:1970gf} that, starting from the 4-point (pt) amplitude at $\ordr (p^2)$, both 6-pt and 8-pt amplitudes could be constructed by recursively imposing the Adler's zero. The relation of such an approach with the current algebra was clarified in Ref.~\cite{Ellis:1970nn}. 

\begin{figure}[t]
\begin{center}
 \includegraphics[width=14cm]{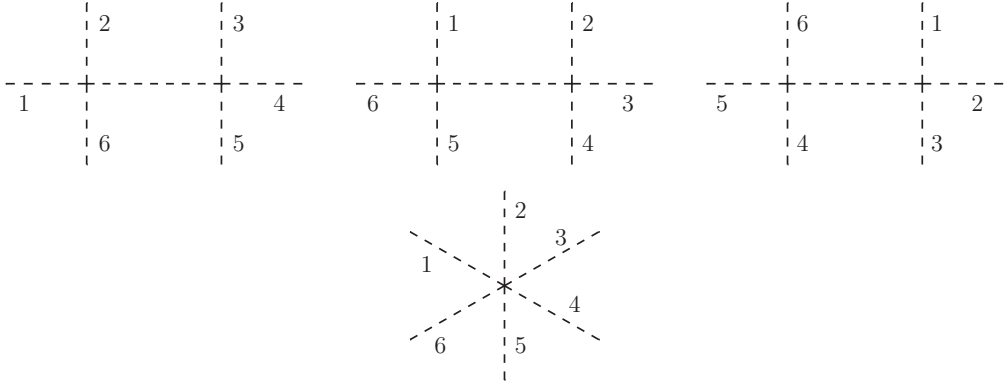}

\end{center}
\caption{\em Feynman diagrams contributing to the 6-pt partial amplitude, $M(1,2,3,4,5,6)$. The top row represents the three factorization channels while the second row is a contact interaction.
}\label{fig:6pt}
\end{figure}

At  first glance it is  surprising that tree amplitudes of pions could be constructed this way, as the Adler's zero condition makes no reference to the $\SU(2)_{\rm L}\times \SU(2)_{\rm R}$ chiral symmetry that is spontaneously broken; the only inputs are IR data. Operationally Susskind and Frye worked with flavor-stripped {\em partial amplitudes}, which is special to pions in the adjoint representation of $\SU(N)$ in that the color factor factorizes simultaneously with the kinematic factorization channel. One then computes the $n$-pt partial amplitude by connecting lower-pt amplitudes through a single internal propagator and summing over all factorization channels. The resulting amplitude does not have the correct soft limit and an $n$-pt contact interaction is added by hand and fully constrained by the Adler's zero condition. Sample Feynman diagrams used in Ref.~\cite{Susskind:1970gf} to bootstrap the 6-pt amplitude are shown in Fig.~\ref{fig:6pt}. Going to higher multiplicities in $n$  makes the procedure quite cumbersome, as  higher-pt contact interactions have to be implemented manually, and little progress was made in the ensuing four decades.

More recently there are two new developments which shed new light on the soft bootstrap program. One is the realization that, for Nambu-Goldstone bosons (NGB's) arising from a global symmetry  $G$  spontaneously broken to a subgroup $H$, the effective Lagrangian depends only on the particular linear representation furnished by NGB's in the unbroken group $H$, and is independent of $G$, up to the normalization of the decay constant $f$ \cite{Low:2014nga,Low:2014oga}. More specifically one imposes a nonlinear shift symmetry in the IR and recursively constructs higher dimensional operators that are invariant under the shift symmetry. The Ward identity of the shift symmetry leads precisely to the Adler's zero condition on S-matrix elements \cite{Kampf:2013vha}, at the leading order in $1/f$. Thus the nonlinear shift symmetry embodies the soft bootstrap program in the Lagrangian approach.

On a separate front, new progresses in the modern S-matrix program lead to on-shell recursion relations for EFT's exhibiting vanishing soft limit \cite{Cheung:2015ota,Cheung:2016drk}, which include NGB's and other more exotic scalar theories. The soft recursion relation allows one to compute the $n$-pt amplitudes directly using factorization channels involving lower-pt sub-amplitudes and sidesteps the need to introduce the $n$-pt contact interaction manually, which greatly streamlines the calculation. The  on-shell soft bootstrap program initially focused on single scalar EFT's, and have since been expanded to supersymmetric theories whose scalar component corresponds to the scalar EFT's with vanishing soft limit \cite{Elvang:2018dco}. Other related works on soft bootstrap can be found in Refs.~\cite{Cheung:2014dqa,Luo:2015tat,Cachazo:2016njl,Bianchi:2016viy,Arkani-Hamed:2016rak,Padilla:2016mno,Rodina:2016jyz,Cheung:2018oki,Bogers:2018zeg,Low:2018acv,Rodina:2018pcb,Yin:2018hht}.

In the context of NGB's,  discussions in the on-shell approach so far concentrate on the leading two-derivative operator in the EFT, naturally. However, the essence of EFT lies in the existence of higher-derivative interactions which become more and more important toward the UV. The higher derivative operator each comes with an incalculable Wilson coefficient encoding the unknown UV physics. In addition, these operators also have more complicated flavor structures that are of multi-trace in nature. It is then interesting to expand the on-shell soft bootstrap program to higher derivative operators and study how these different aspects of EFT's arise from the IR, which is the aim of this work.

More broadly, studying higher derivative operators from the on-shell perspective could have far reaching implications on several other fascinating aspects of modern S-matrix program. One is the ``double-copy'' structure \cite{Bern:2008qj} that is prevalent among many quantum field theories, including  scalar EFT's \cite{Du:2016tbc,Carrasco:2016ldy,Cheung:2016prv,Carrasco:2016ygv,Cheung:2017yef,Mizera:2018jbh,Bjerrum-Bohr:2018jqe}. This structure is manifest in the Cachazo-He-Yuan (CHY) representation of scattering amplitudes for massless particles \cite{Cachazo:2013hca,Cachazo:2013iea}. However, most of the studies so far are confined to the leading operator in the derivative expansion, except for Refs.~\cite{He:2016iqi,Carballo-Rubio:2018bmu} which considered higher derivative operators in gauge theories and gravity. For the NLSM, work has only been done in the special cases when the higher order corrections satisfy certain properties, such as Bern-Carrasco-Johansson (BCJ) relations \cite{Carrasco:2016ldy,Elvang:2018dco} or subleading double soft theorems \cite{Rodina:2018pcb}. Part of the purpose of this work is to initiate a study on the most general higher derivative operators in scalar EFT's in the on-shell approach.

An interesting aspect of higher derivative operators is that often they involve color/flavor factors that are of multi-trace in nature. Most quantum field theories studied by the ``scattering amplitudes'' community currently involve fields that carry no color/flavor charges or transforming under the adjoint representation of $\SU(N)$ color/flavor group. This is due to the U(1) decoupling relation for the adjoint of $\SU (N)$ theory \cite{Elvang:2013cua}, which allows the simultaneous factorization of color/flavor and kinematic factors in the factorization channel. As a consequence, the relation between the  full amplitudes and the color/flavor-stripped partial amplitudes is simple.  This is not true for a general representation of a classical Lie group, and the adjoint of $\SU(N)$ is the only known example so far. By studying the multi-trace property of higher derivative operators, we uncover a new possibility exhibiting the same simultaneous factorization of color/flavor and kinematic factors, which involves the {\em fundamental} representation of $\SO (N)$ group. The new example enjoys the same simple relation between the full and partial amplitudes as in the adjoint of $\SU (N)$. This opens a door to study whether any of the fascinating features involving the $\SU (N)$ would persist for the fundamental of $\SO (N)$, which is outside of the scope of current work.

This work is organized as follows. In Section \ref{sect:leading} we first consider soft-bootstrapping the leading two-derivative operator in EFT's with vanishing soft limit, with a focus on multi-scalar EFT's and pointing out new subtleties involved. We introduce the notion of a ``soft block'' here, which serves as the seed of soft bootstrap, and present a new EFT with a flavor structure that is different from the commonly studied $\SU (N)$ adjoint representation. In Section \ref{sect:higherD} we introduce soft blocks at the four-derivative order, which include both the parity-even and the parity-odd soft blocks, and study their soft-bootstrap. The parity-odd soft block obviously maps to the Wess-Zumino-Witten term in the NLSM. Each soft block at this order comes with an undetermined free parameter, which is later shown to correspond to the Wilson coefficient in the EFT language. It is also in this section where we demonstrate the existence of two consistent EFT's in soft bootstrap. Then in Section \ref{sec:mat} we explicitly match these two EFT's to NLSM's based on the coset $\SU (N)\times \SU (N)/ \SU (N)$ and $\SO (N+1)/\SO (N)$, which is followed by the summary and outlook. We also provide two appendices on multi-trace flavor-ordered partial amplitudes and the IR construction of NLSM effective Lagrangians.

\section{Leading Order Soft Bootstrap}
\label{sect:leading}

\subsection{An Overview}
\label{sec:sbo}
The modern approach to the soft bootstrap of  EFT's was initiated in Refs.~\cite{Cheung:2015ota,Cheung:2016drk}, which considered various kinds of one-parameter scalar effective theories. Much of the discussion there focused on single scalar EFTs. Since our main focus is NLSMs with multiple scalars, here we give an overview of soft bootstrap procedure adapted to multi-scalar EFTs.

We use the notation ${\Phi}= \{\phi_1, \phi_2,\cdots\}$ to denote a generic set of scalars. By construction an EFT consists of an infinite number of operators organized with increasing powers of derivatives and fields. In general the EFT is a double expansion in the two  parameters:
\be
\frac{\partial^\mu}{\Lambda} \qquad \text{and} \qquad \frac{{\Phi}}{f} \ ,
\ee
where $\Lambda$ and $f$ are two mass scales which characterize the derivative expansion and the field expansion, respectively. Then the effective Lagrangian will have the schematic form 
\be
\label{eq:Lmultiscalar}
{\cal L}_\Phi= \Lambda^2 f^2 \sum_{m\ge 1; n\ge 2} \frac{c_{2m,n}}{\Lambda^{2m} f^{n}} \partial^{2m} \left[{\Phi}\right]^{n} \ ,
\ee
where we have suppressed the Lorentz indices on the derivatives and Lorentz invariance implies an even number of derivatives. In addition $\left[{\Phi}\right]^{n}$ denotes generic contractions of $n$ $\phi_i$ scalars. So at a given $(2m,n)$ there could be many Wilson coefficients $c_{2m,n}$. The overall factor $\Lambda^2f^2$ is dictated by a canonically normalized kinetic term for $(\partial \phi)^2$, which also requires $c_{2,2}=1$.

A prime example of EFTs in the soft bootstrap program is the leading two-derivative operator in $\SU (N) \times\SU (N) / \SU (N)$ NLSM, which describes a set of massless scalars transforming under the adjoint representation of $\SU (N)$. One major advantage of working with $\SU (N)$ NLSM is the existence\ of flavor-ordered partial amplitudes with a simple factorization property \cite{Kampf:2013vha}, much like the color-ordered partial amplitudes in the $\SU (N)$ Yang-Mills theory. The full amplitude at order ${\cal O}(p^2)$ can be written as
\bea
M^{a_1 \cdots a_{n} } (p_1, \cdots, p_{n}) \equiv \sum_{\sigma \in S_{n-1}} \mathcal{C}^{ a_{\sigma (1)} \cdots a_{\sigma (n-1)} a_n} M( \sigma(1), \cdots, \sigma (n-1), n )\ ,\label{eq:stfod}
\eea
where 
\bea
\mathcal{C}^{a_1  a_2 \cdots a_n } = \tr \left( T^{a_1} T^{a_2} \cdots T^{ a_n} \right)\ , \label{eq:ffat}
\eea  
 is the flavor factor, $\sigma$ is a permutation of indices $\{1, 2, \cdots, n-1\}$ and $T^a$ is the generator of $\SU (N)$ group.  The full amplitude is permutation invariant among all external legs, while the flavor-ordered partial amplitude $M(1,2,\cdots n)$ is  invariant only under {\em cyclic} permutations of the external legs because of the cyclic property of the trace in Eq.~(\ref{eq:ffat}). 
 
\begin{figure}[t]
\centering
\includegraphics[width=12.13cm]{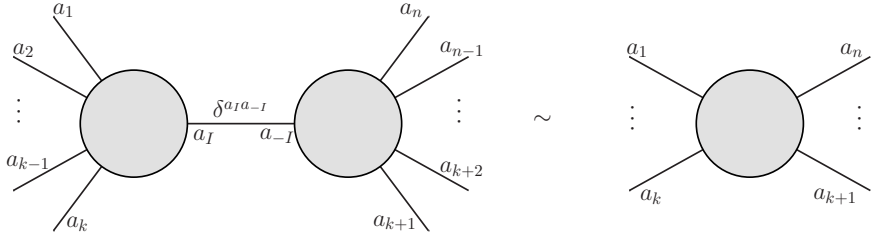}
\caption{ {\em The merging of flavor factors.}\label{fig:mergt}}
\end{figure}
 
 At the two-derivative level, the flavor factor $\mathcal{C}^{a_1  a_2 \cdots a_n }$ can always be written as a {\em single} trace operator involving  $T^a$.\footnote{This is a general statement independent of the coset $G/H$, as the scalars are always in the adjoint representation of $G$.} Furthermore, the $\SU (N)$ generators satisfy the following completeness relation:
 \bea
\label{eq:suNcompl}
&& \sum_{a_I}\ \tr \left( T^{ a_1}  \cdots T^{  a_k} T^{ a_I} \right) \times \tr \left( T^{ a_I} T^{ a_{k+1}}  \cdots T^{  a_n} \right) =  \nonumber\\
&&\qquad  \tr \left( T^{ a_1} T^{ a_2} \cdots T^{  a_n} \right)  - \frac1{N}  \tr \left( T^{ a_1}  \cdots T^{  a_{k-1}} \right)\times  \tr \left( T^{ a_k}  \cdots T^{  a_n} \right)\ .
\eea
In a tree diagram the disconnected $1/N$ term does not contribute due to the decoupling relations of the $\U (1)$ Goldstone boson in $\U (N)\times \U (N)/\U(N)$ NLSM. (See, for example, Ref.~\cite{Elvang:2013cua}.) This is an important property for the on-shell recursion relation, which expresses a higher-pt partial amplitude as the product of two or more lower-pt sub-partial amplitudes. When dressing up each sub-partial amplitude with its own flavor factor, the recursed higher-pt amplitude now has the following flavor factor:
\be
\sum_{ a_I}\ \mathcal{C}^{ a_1  a_2 \cdots  a_k  a_I} \times \mathcal{C}^{ a_I  a_{k+1}  a_{k+2}  \cdots  a_n} \sim \mathcal{C}^{ a_1  a_2 \cdots  a_n} \ ,
\ee
which is a single trace flavor factor that preserves the ordering of sub amplitudes. Note the summed-over index $ a_I$ arises from the internal propagator $i\delta^{ab}/p^2$, as shown in Fig.~\ref{fig:mergt}. This connection between soft recursion of partial amplitudes and proper factorization of flavor factors is very special for the adjoint representation of $\SU (N)$ group, and the main reason why so far most studies on NLSM utilizing partial amplitudes  assume the field content to transform as the adjoint of $\SU (N)$. For example, generators in the adjoint of $\SO (N)$ group do not have as nice a completeness relation as in Eq.~(\ref{eq:suNcompl}), and the corresponding partial amplitudes do not enjoy the simple factorization property. However, later we will see a new possibility to make compatible the partial amplitudes with factorization using multi-trace operators, which is matched to the $\SO(N+1)/\SO(N)$ coset, where the NGB's transform as the {\em fundamental} representation of $\SO(N)$.

Given the partial amplitudes, the soft bootstrap program then constructs higher-pt amplitudes from lower-pt amplitudes by using  the soft recursion relation \cite{Cheung:2015ota}, which utilizes an all-leg shift for the external momenta of an $n$-pt amplitude $M_n$,
\bea
\label{eq:alllegshift}
p_i \to \hatp_i = (1-a_i z)\, p_i\ ,
\eea
where $z$ is a complex shift parameter and total momentum conservation requires, choosing all momenta to be incoming,
\bea
\sum_{i=1}^n a_i\, p_i^{\mu} = 0\label{eq:smcc} \ .
\eea
The deformed amplitude $\hatm_n (z)$ with momentum variables $\hatp_i$ is still an on-shell amplitude, in the sense that all external momenta remain on-shell, and taking the soft limit of $p_i$ corresponds to setting $z \to 1/a_i$. 

In $D$-dimensional spacetime Eq.~(\ref{eq:smcc}) can be viewed as a set of $D$ linear equations in $n$ variables $a_i$. Because of momentum conservation $\sum_i p_i=0$, there are really only $n-1$ generic momenta and a trivial solution where all $a_i$'s are equal always exists. When $n-1 \ge D+1$, these generic momenta become linearly dependent and non-trivial solutions exist. The number of distinct non-trivial solutions is $n-D-1$, which is $n-5$ in $D=4$. Furthermore, when a solution exists, rescaling all $a_i$'s simultaneously lead to a ``degenerate'' solution. In the end, the general solution for $a_i$'s can be written as
\be
\label{eq:aigen}
\{a_i\} = \sum_{r=1}^{n-5} A^{(r)} \{a_i^{(r)}\} +B \ ,
\ee
where $\{a_i^{(r)}\}$, $r=1, \cdots, n-5$, are the non-trivial solutions which can be expressed in terms of kinematic invariants of external momenta, while $A^{(r)}$ and $B$ are arbitrary constants  reflecting the re-scaling degrees of freedom in both the non-trivial and trivial solutions. It is rather intriguing that the solutions for $a_i$'s have a ``shift symmetry'' and are  defined ``projectively.''

Since the Adler's zero condition requires  the amplitude in NLSM  to vanish linearly in the soft momentum $\hatm_n(z)\sim \hat{p}$ for $\hat{p}\to 0$  \cite{Adler:1964um}, one can define a soft factor,
\bea
F_n(z) \equiv \prod_{i=1}^n (1-a_i z)\ ,\label{eq:sffn}
\eea
so that the Cauchy integral below always vanishes,
\bea
\oint \frac{dz}{z} \frac{\hatm_n (z) }{F_n(z)}=0\ . \label{eq:cauc}
\eea
This is because the integrand at large $z$ vanishes like $\ordr (z^{1-n})$ and the residue at the infinity is zero. As a consequence of the Adler's zero condition, the integrand has no poles at $z = 1/a_i$ and the only poles come from $z = 0$ as well as the factorization channel $I$, the shifted internal momentum of the corresponding propagator being $\hatP_I(z)= P_I+z Q_I $, where
\be
P_I = \sum_{i \in I} p_i \qquad \text{and}\qquad Q_I =- \sum_{i\in I} a_i \, p_i \ .
\ee
 The residue theorem then relates the residue at $z=0$, which is nothing but the $n$-pt amplitude $M_n$, to the other residues at $\hatP_I^2(z_I^\pm)=0$:
\bea
M_n = \hatm_n (0) = -\sum_{I, \pm} \frac1{P_I^2} \frac{\hatm_L^{(I)}  (z_I^\pm) \hatm_R^{(I)}  (z_I^\pm) }{F_n(z_I^\pm) (1-z_I^\pm/z_I^\mp )}\ , \label{eq:sbrr}
\eea
where $M_L$ and $M_R$ are the two lower-pt on-shell amplitudes associated with the factorization channel $I$. Therefore, starting with some ``seed'' amplitudes that does not factorize, one can use Eq.~(\ref{eq:sbrr}) to recursively construct on-shell amplitudes of all multiplicities in the theory. We will refer to Eq.~(\ref{eq:sbrr}) as the soft recursion relation.

A   useful  special case  is when all $M_L^{(I)}$ and $M_R^{(I)}$ are local functions of momenta, i.e. without poles from any factorization channels. It is based on the observation that the contribution from a particular factorization channel $I$ in Eq.~(\ref{eq:sbrr}) is the residue at $z=z_I^\pm$ of the analytic function
\be
\label{eq:localf}
 \frac{\hatm_L^{(I)} (z) \hatm_R^{(I)} (z) }{z F_n(z) \hat{P^2_I} (z)} \ ,
\ee
which has poles at $z=0$ and $z=1/a_i$. The poles at $z=1/a_i$ comes about because the individual factorization channel $I$ does not satisfy Adler's zero condition in the soft limit; only after summing over all factorization channels is the Adler's condition satisfied and the $1/a_i$ poles disappear. Then a second application of the residue theorem relates the residues of Eq.~(\ref{eq:localf}) at $a=z_I^\pm$ to the residues at $z=0, 1/a_i$:
\bea
M_n = -\sum_I \left[ \frac{M_L^{(I)} M_R^{(I)}}{P_I^2} + \sum_{i=1}^n \Res_{z=1/a_i} \frac{\hatm_L^{(I)} (z) \hatm_R^{(I)} (z) }{z F_n(z) \hat{P^2_I} (z)} \right]\ .\label{eq:sbrrl}
\eea
The first term on the right-hand side represents all Feynman diagrams contributing to $M_n$ that contain an internal propagator. The second term, on the other hand, must then be a local function of external momenta and relates directly to the $n$-pt contact operator in the effective Lagrangian.

The soft bootstrap program is predictive only when the higher-pt amplitudes constructed using the soft recursion relation are independent of the arbitrary coefficients $A^{(r)}$ and $B$ in the general solution of $a_i$'s in Eq.~(\ref{eq:aigen}). Otherwise we would introduce more and more unknown parameters as we go to higher-pt amplitudes. Therefore, we define a consistent EFT in soft bootstrap to be when
\begin{itemize}
\item[] \em The amplitude $M_n$ obtained from the soft recursion relation is independent of the arbitrary constants $A^{(r)}$ and $B$ for all $n$. 
\end{itemize}
Otherwise the EFT one is trying to construct using soft bootstrap simply does not exist.

\subsection{Introducing the Soft Blocks}
\label{subsect:softblocks}

At this point it is convenient to introduce the notion of a ``soft block,'' 
\begin{itemize}
\item A soft block ${\cal S}^{(k)}(p_1,\cdots, p_n)$ is a contact interaction carrying $n$ scalars and $k$ derivatives that satisfies the Adler's zero condition when all external legs are on-shell. 
\end{itemize}
Because the soft blocks themselves satisfy the Adler's zero condition, they can be used as a seed amplitude in the recursion relation. As such, the soft block is an input to soft bootstrap. 

For $k \le 4$, which we focus on in this work, the soft blocks exist only for $n=4$ and $n=5$. It cannot exist for $n=3$ because there is no non-trivial kinematic invariant built out of three on-shell real momenta satisfying total momentum conservation. Beyond $n=3$, let's perform an all-leg-shift as in Eq.~(\ref{eq:alllegshift}) on the external momenta. The ``shifted block'' is now a polynomial of degree $k$ in $z$: $\hat{\cal S}^{(k)}=\hat{\cal S}^{(k)}(z)$. However, in $D=4$  there exists non-trivial solutions for $a_i$'s only when the number of external legs $n \ge 6$. Since $\hat{\cal S}_n^{(k)}(z)$ satisfies the Adler's zero condition by assumption, it must have
\be
\hat{\cal S}^{(k)}(1/a_i)=0, \qquad i=1, \cdots, n\ , \quad n\ge 6 \ .
\ee
In other words $\hat{\cal S}_n^{(k)}(z)$ should have $n$ distinct roots in $z=1/a_i$, which cannot happen for a polynomial of degree $k\le 4$. Therefore, at four-derivative order or less, a soft block can exist only if it contains $n\le 5$ external momenta.

\label{sec:2dsb}

\subsection{Single-trace Soft Block at ${\cal O}(p^2)$}
\label{sec:op2so}

\begin{figure}[t]
\centering
\includegraphics[width=4cm]{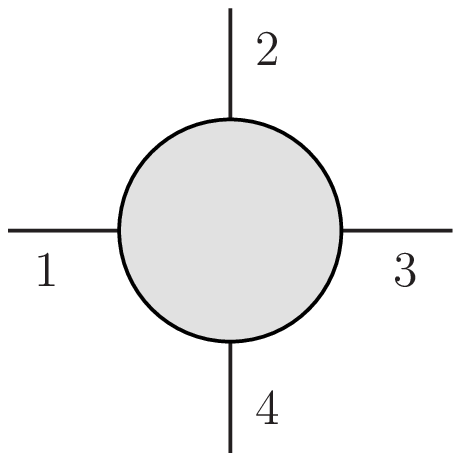}\qquad \includegraphics[width=4cm]{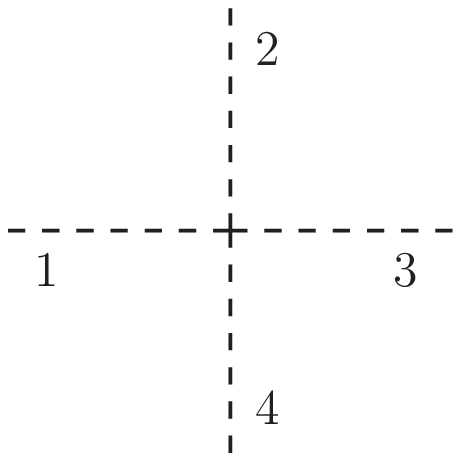}
\caption{\em The 4-pt single-trace soft block ${\cal S} (1,2,3,4)$, as well as the corresponding Feynman vertex. We use the convention that a solid line represents an on-shell scalar particle, while a dashed line represents an off-shell scalar. \label{fig:sb4s}}
\end{figure}

Next we identify soft blocks at ${\cal O}(p^2)$ that are invariant under cyclic permutations of all external legs, which we refer to as the single-trace soft block.  Working with partial amplitudes, we start with the 4-pt flavor-ordered soft block ${\cal S}^{(2)}(1,2,3,4)$  such that 
\begin{enumerate}
\item ${\cal S}^{(2)}(1,2,3,4)$ is quadratic in external momenta. 
\item ${\cal S}^{(2)}(1,2,3,4)$ satisfies the Adler's zero condition.
\item ${\cal S}^{(2)}(1,2,3,4)$ is invariant under cyclic permutations of all external legs.
\end{enumerate}
At 4-pt level there are only two independent kinematic invariants $s_{12}$ and $s_{13}$, where  $s_{ij} \equiv (p_i +p_j)^2$. Writing down the most general kinematic invariant and imposing the second and the third conditions lead to a unique soft block, up to total momentum conservation, 
\bea
{\cal S}^{(2)}(1,2,3,4) &=& c_0\, \frac{s_{13}}{f^2}\ ,\label{eq:o2sos}
\eea
where $c_0$ is a constant parameter. Using momentum conservation one could rewrite the right-hand side as either $s_{24}$, thereby exhibiting the cyclic property. This soft block is presented  in Fig. \ref{fig:sb4s} and corresponds to a two-derivative operator of the form
\be
\label{eq:softope}
\frac{1}{f^2} \partial^2 \left[ \Phi\right]^4 
\ee
in the effective Lagrangian in Eq.~(\ref{eq:Lmultiscalar}). 

One could ask if there is a 5-pt soft block at the two-derivative level. In this case there are 5 independent kinematic invariants, $\{s_{12}, s_{23}, s_{34}, s_{45}, s_{51}\}$. It is simple to check that no linear combination of these five invariants could satisfy conditions 2 and 3 simultaneously.

\begin{figure}[t]
\begin{center}
 \includegraphics[width=14cm]{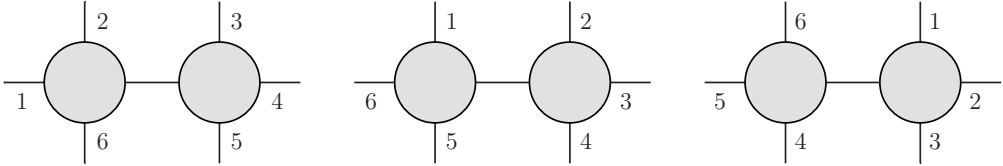}
\end{center}
\caption{\em The three factorized soft blocks contributing to  the single trace 6-pt amplitude $M(1,2,3,4,5,6)$.
}\label{fig:st6pfc}
\end{figure}



Starting with ${\cal S}^{(2)}(1,2,3,4)$, we soft-bootstrap the 6-pt partial amplitude by Eq. (\ref{eq:sbrrl}):
\bea
\label{eq:single6pt}
M^{(2)} (1,2,3,4,5,6) &=& -\frac{c_0^2}{f^4} \left[ \frac{s_{13} s_{46}}{P^2_{123}} + \frac{s_{24} s_{15}}{P^2_{234}} +\frac{ s_{35} s_{26}}{P^2_{345}} - P^2_{135}\right],
\eea
where $P_{i_1 i_2 \cdots i_n} \equiv p_{i_1} + p_{i_2} + \cdots + p_{i_n}$. This is a consistent amplitude because it does not depend on $A^{(r)}$ and $B$ in the general solution for $a_i$'s. 

In soft-bootstrap the 6-pt amplitude is given by three factorized soft blocks shown in Fig.~\ref{fig:st6pfc}. Here we wish to make a distinction between ``factorized soft blocks,'' which contribute to the right-hand side of soft recursion in Eq.~(\ref{eq:sbrr}), and the Feynman diagrams which may contain non-factorizable contact terms. More explicitly, we show in Fig.~\ref{fig:6pt} the Feynman diagrams contributing to the 6-pt amplitude, which have three factorizable diagrams and one non-factorizable 6-pt contact term.  Contributions from the three factorized soft blocks in Fig.~\ref{fig:st6pfc} is equal to the four Feynman diagrams in Fig.~\ref{fig:6pt}.

The 6-pt contact interaction bootstrapped from the soft block is of the form
\be
\frac1{f^4} \partial^2 \left[\Phi\right]^6 \ ,
\ee 
whose presence makes the 6-pt amplitude conform to the Adler's zero condition. At the Lagrangian level,  the structure and the coefficient of this operator is constrained by a shift symmetry in the IR construction of effective Lagrangians \cite{Low:2014nga,Low:2014oga}. Once the 6-pt amplitude is soft-bootstrapped, one then proceeds to higher-pt amplitudes using in Eq.~(\ref{eq:sbrr}).

The preceding discussion on the 6-pt amplitude leads to the question: what operators  in the effective Lagrangian in Eq.~(\ref{eq:Lmultiscalar}) can be  soft-bootstrapped from ${\cal S}^{(2)}(1,2,3,4)$?

To answer this question, we need to digress a little bit and recall some definitions to characterize the property of a Feynman diagram. Let's use $b_i$ and $d_i$ to represent the number of scalars and derivatives, respectively, carried by a particular operator ${\cal O}_i$ in Eq.~(\ref{eq:Lmultiscalar}). Generically a tree Feynman diagram with $n$ external legs can be expressed as a rational function of degree $d$ in external momenta. Then $n$ and $d$ are given as
\bea
\label{eq:nequation}
n &=&\sum_i n_i \, b_i-2 I_B \ ,\\
\label{eq:dequation}
d &=& \sum_i n_i\, d_i -2 I_B \ ,
\eea
where $I_B$ is the number of internal propagators and $n_i$ is the number of insertions of ${\cal O}_i$ vertex in the diagram. Eq.~(\ref{eq:nequation}) comes from conservation of the number of scalar fields, while Eq.~(\ref{eq:dequation})  is simply subtracting the power of momentum in the denominator from the numerator in the diagram. In a general Feynman diagram, the number of loops $L$ is given by
\be
L= I_B - \sum_i n_i +1 \ ,
\ee
which comes about by counting the number of unconstrained momenta in a diagram: each internal propagator has a momentum integral and each insertion of vertex has a momentum delta function, and one delta function  simply  enforces total momentum conservation. Setting $L=0$ to replace $I_B$ in Eqs.~(\ref{eq:nequation}) and (\ref{eq:dequation}) we have
\bea
\label{eq:nequationnew}
n &=&\sum_i n_i \, (b_i-2) +2 \ ,\\
\label{eq:dequationnew}
d &=& \sum_i n_i\, (d_i-2) +2 \ .
\eea
If all  operators in Eq.~(\ref{eq:Lmultiscalar}) are such that 
\be
\rho \equiv \frac{d_i-2}{b_i-2}
\ee
is a fixed non-negative rational number, then every Feynman diagram in the EFT will also have $(d-2)/(n-2)=\rho$, as can be seen from plugging $\rho$ into Eqs.~(\ref{eq:nequationnew}) and (\ref{eq:dequationnew}). In this sense operators carrying a definitive $\rho$ form a closed set among themselves. The parameter $\rho$ was first introduced in Refs.~\cite{Cheung:2015ota,Cheung:2016drk} to characterize a particular power counting order in derivative.\footnote{Ref.~\cite{Elvang:2018dco} introduced the quantity $\tilde{\Delta}$, which generalizes $\rho$ to external states with spins. For pure scalar theories, $\tilde{\Delta}=\rho+1$.} 

The operator corresponding to the soft block ${\cal S}^{(2)}(1,2,3,4)$ has $n_i=4$, $d_i=2$, and  $\rho=0$. Thus all operators with $\rho=0$ can be soft-bootstrapped from this particular soft block. This can also be seen explicitly from Eq.~(\ref{eq:dequationnew}): an arbitrary number of insertions of the two-derivative soft block will generate an amplitude with two powers of external momenta. In addition, since we start with a soft block with $n_i=4$, Eq.~(\ref{eq:nequationnew}) shows the number of external legs $n$ must also be an even number. We conclude that the Wilson coefficients of operators of the form
\be
\label{eq:multifk}
\frac1{f^{2k-2}} \partial^2 \left[\Phi\right]^{2k} \ , \qquad k > 2\ ,
\ee 
are completely determined by soft bootstrap. The only free parameter is the unknown coefficient $c_0$ in Eq.~(\ref{eq:o2sos}), which can be absorbed into the definition of $f$. This agrees with the outcome of imposing shift symmetries in the Lagrangian to bootstrap two-derivative operators in NLSMs that are higher orders in the $1/f$ expansion \cite{Low:2014nga,Low:2014oga}. In fact, in the Lagrangian approach the two-derivative operators can be resummed to all orders in $1/f$ into a simple compact form that is invariant under the shift symmetry, which is briefly reviewed in Appendix \ref{app:allop}.  The EFT constructed from the single trace soft block in Eq.~(\ref{eq:o2sos}) is  the  leading two-derivative operator in the familiar $\SU (N)$ NLSM.

It is worth commenting early on that a power counting in terms of $\rho$ is of limited use for multi-scalar EFTs, especially when going beyond leading two-derivative order in NLSMs. In particular, ${\cal O}(p^4)$ operators in NLSMs involve an infinite number of operators carrying four derivatives but an arbitrary number of scalar fields, as we will see later. The reason a power counting based on $\rho$ is useful for single scalar EFT's, at least for the leading interactions, is because there is a unique two-derivative operator that is the kinetic term. In fact, in single scalar EFT's all two derivative operators of the form,
\be
\label{eq:singleSca}
\frac{\phi^k}{f^k} \partial_\mu \phi \partial^\mu \phi \ ,
\ee
can be removed by a field re-definition, $\phi\to \phi+ F(\phi)$, for a suitably chosen $F(\phi)$.  Equivalently, tree amplitudes of $n\ge 3$ identical massless scalars must vanish at ${\cal O}(p^2)$. This can be seen easily because  Bose symmetry requires the  amplitude must be completely symmetric in interchange of any two external momenta,
\be
\label{eq:Msinglef}
M(p_1, \cdots, p_{n}) \propto (p_1+\cdots+p_{n})^2 = 0 \ ,
\ee 
which vanishes due to total momentum conservation. However, when there is more than one ``flavor'' of massless scalar, Bose symmetry only requires a symmetric ``wave function'' under the simultaneous exchange of momentum {\em and} flavor quantum numbers. Therefore, operators in Eq.~(\ref {eq:multifk}) do have non-zero S-matrix elements starting at ${\cal O}(p^2)$ and their coefficients are determined by soft bootstrap. This means that a direct power counting of derivatives, or $\Lambda$, suffice.\footnote{One can force the $\ordr (p^2)$ interactions to vanish for multi-scalar EFTs, as in multi-field DBI \cite{Cheung:2016drk}. For the leading order interactions of such a theory, the recognition of a fixed $\rho$ is still useful.} When going to higher order corrections, the $\ordr (p^2)$ interactions feed back into the soft bootstrap of ${\cal O}(p^4)$ vertices, and $\rho$ is no longer fixed at this order.

\subsection{Double-trace Soft Block at ${\cal O}(p^2)$}
\label{sec:op2doub}

\begin{figure}[t]
\centering
\includegraphics[width=4cm]{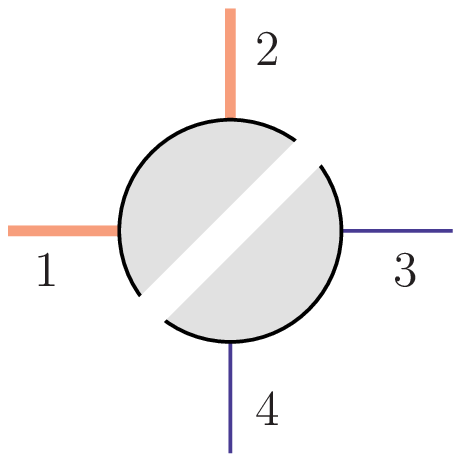}\qquad \includegraphics[width=4cm]{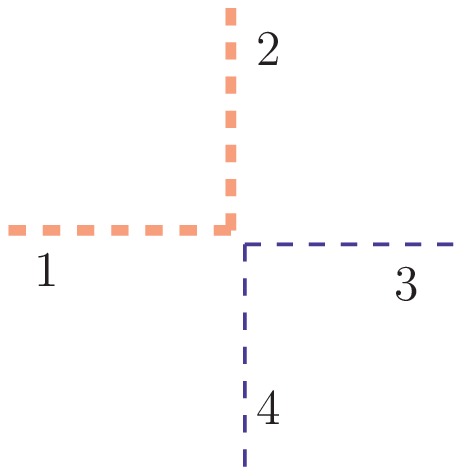}
\caption{\em The 4-pt double-trace soft block ${\cal S} (1,2|3,4)$, as well as the corresponding Feynman vertex. Notice the two separate flavor flows of $(1,2)$ and $(3,4)$.\label{fig:sb4d}}
\end{figure}

In the previous subsection we constructed a soft block that is invariant under cyclic permutations of {\em all} external legs. One could ask if this assumption can be relaxed. Indeed here we consider a new soft block ${\cal S}^{(2)}(1,2|3,4)$ that satisfies the first two requirements in Section \ref{sec:op2so} and 
\begin{itemize}
\item ${\cal S}^{(2)}(1,2|3,4)$ is invariant under separate cyclic permutations of $(1,2)$ and $(3,4)$.
\end{itemize}
We do not consider a soft block that is invariant under only cyclic permutations of three external legs such as ${\cal S}^{(2)}(1,2,3|4)$, which would imply the soft block is not neutral under the flavor charge. We call ${\cal S}^{(2)}(1,2|3,4)$ the ``double trace'' soft block, which is given by
\be
{\cal S}^{(2)} (1,2|3,4) = \frac{d_0}{f^2} s_{12}\ ,\label{eq:p2dtsb}
\ee
up to total momentum conservation. Diagrammatically we present the double trace soft block as in Fig. \ref{fig:sb4d}. Similar to the single trace case, we do not find any 5-pt soft blocks that are quadratic in external momenta.

\begin{figure}[t]
\centering
\includegraphics[width=14cm]{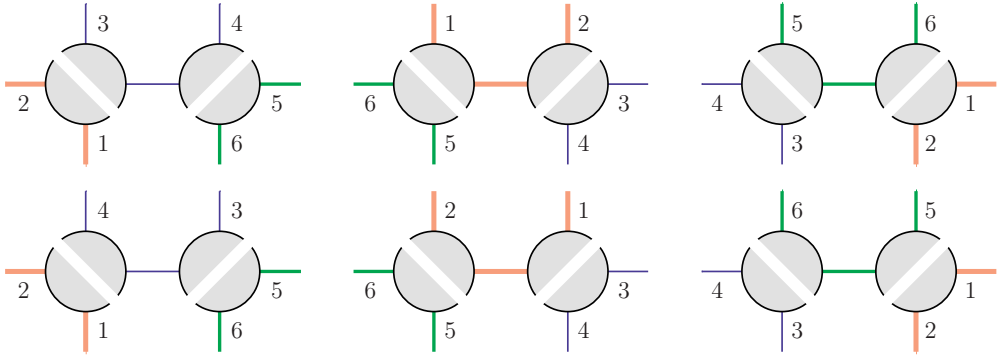}
\caption{\em The 6 factorization channels of $M (1,2|3,4|5,6)$. In each column we symmetrize over the flavor indices that are connected 
through the internal propagator. \label{fig:sb6d}}
\end{figure}

\begin{figure}[t]
\centering
\includegraphics[width=9.7cm]{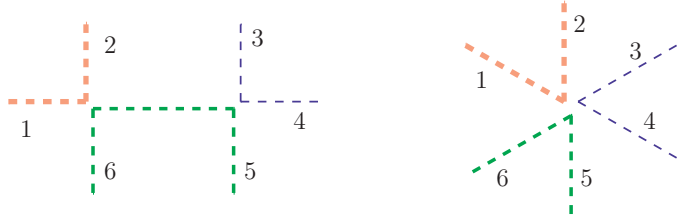}
\caption{\em Two classes of Feynman diagrams of $M (1,2|3,4|5,6)$. The diagram on the left contains a propagator, and there are 6 diagrams of this kind, corresponding to the 6 factorization channels. The diagram on the right is the contact term.\label{fig:fd6dou}}
\end{figure}

Using ${\cal S}^{(2)} (1,2|3,4)$ we  can construct  6-pt amplitudes with different flavor-orderings. For example, using Eq.~(\ref{eq:sbrrl}) we obtain the partial amplitude $M^{(2)} (1,2|3,4|5,6)$ that is invariant under three separate (cyclic) permutations in $(1,2), (3,4)$ and $(5,6)$. There are six factorized soft blocks shown in Fig.~\ref{fig:sb6d}. It is important to draw a contrast with the three factorized blocks, in the case of single trace soft block, shown in Fig.~\ref{fig:st6pfc}. The additional factorized blocks, shown in the second row of Fig.~\ref{fig:sb6d}, come about because we need to symmetrize with respect to the flavor indices $(ij)$ that are connected through the internal propagator, so as to make the amplitude invariant under the cyclic permutation $(ij)\to (ji)$. In fact, if we only included the three factorized blocks in Fig.~\ref{fig:st6pfc} for the double-trace soft blocks, the $a_i$'s dependence would not cancel and the resulting amplitude is inconsistent. Only after summing over all six contributions in Fig.~\ref{fig:sb6d} did we arrive at a consistent amplitude,
\bea
\label{eq:m2123456}
M^{(2)} (1,2|3,4|5,6) &=& -\frac{d_0^2}{f^4} \left[ s_{12} s_{56} \left(\frac{1}{P^2_{124}} + \frac{1}{P^2_{123}} \right) + s_{12} s_{34} \left(\frac{1}{P^2_{125}} + \frac{1}{P^2_{126}} \right) \right.\non\\
&&\left.+ s_{34} s_{56} \left(\frac{1}{P^2_{134}} + \frac{1}{P^2_{234}} \right)  \right] + M^{(2),\ctt} (1,2|3,4|5,6),
\eea
where the contact term $M^{(2),\ctt}$ is given by
\bea
M^{(2),\ctt} (1,2|3,4|5,6) &=& -\frac{d_0^2}{f^4}\sum_{i=1}^6 \Res_{z=1/a_i} \ \frac{1 }{z F_6(z) } \left[ \hats_{12} \hats_{56} \left(\frac{1}{\hatP^2_{124}} + \frac{1}{\hatP^2_{123}} \right)  \right.\non\\
&&\left.+ \hats_{12} \hats_{34} \left(\frac{1}{\hatP^2_{125}} + \frac{1}{\hatP^2_{126}} \right) + \hats_{34} \hats_{56} \left(\frac{1}{\hatP^2_{134}} + \frac{1}{\hatP^2_{234}} \right)  \right]\non\\
&=&-\frac{d_0^2}{f^4}\sum_{i=1}^6 \Res_{z=1/a_i} \frac{1 }{z F_6(z) } \left( \hats_{12} +\hats_{34} +\hats_{56}\right)\non\\
&=&  \frac{d_0^2}{f^4}\ \Res_{z=0} \frac{1 }{z F_6(z) } \left( \hats_{12} +\hats_{34} +\hats_{56}\right)\non\\
&=& \frac{d_0^2}{f^4}\ \left( s_{12} + s_{34} + s_{56} \right).
\eea
Therefore, this particular 6-pt amplitude is
\bea
\label{eq:sonp26pa}
M^{(2)} (1,2|3,4|5,6) &=& -\frac{d_0^2}{f^4} \left[ s_{12} s_{56} \left(\frac{1}{P^2_{124}} + \frac{1}{P^2_{123}} \right) + s_{12} s_{34} \left(\frac{1}{P^2_{125}} + \frac{1}{P^2_{126}} \right) \right.\non\\
&&\left.+ s_{34} s_{56} \left(\frac{1}{P^2_{134}} + \frac{1}{P^2_{234}} \right) - s_{12} - s_{34} - s_{56} \right]\ ,
\eea
which is manifestly invariant under the three separate cyclic permutations and satisfies the Adler's zero condition. It is interesting that requiring the recursed amplitude to be independent of  $a_i$  forces a flavor structure that is of the ``triple trace'' nature.

Eq.~(\ref{eq:m2123456}) is  a different amplitude  from the 6-pt amplitude in Eq.~(\ref{eq:single6pt}), which is bootstrapped from the single trace soft block. The corresponding Feynman diagrams are shown in Fig. \ref{fig:fd6dou}, which include the factorization channels as well as the 6-pt contact interaction.

In computing the 6-pt amplitude we do not have to explicitly plug in the solutions for $a_i$ because of the specialized recursion relation in Eq.~(\ref{eq:sbrrl}). When going to 8-pt amplitude this is not true anymore and one need to check whether the 8-pt amplitude is consistent in that it doesn't depend on the arbitrary coefficients $A^{(r)}$ and $B$ defined in Eq.~(\ref{eq:aigen}). We have checked numerically that this is indeed the case.

The EFT constructed from the double trace soft block is  a new theory different from the single trace soft block, which corresponds to the $\SU(N) \times \SU(N)/\SU(N)$ NLSM, with the massless scalars transforming under the adjoint representation of  $\SU(N)$.  We will see that the new EFT corresponds to $\SO(N+1)/\SO(N)$ NLSM, where $N$ massless scalars transform under the fundamental representation of $\SO (N)$ group.\footnote{A special case, the $\SO (3)/ \SO (2)$ NLSM was constructed in Ref. \cite{Cheung:2016drk} using soft bootstrap, by starting with 2 flavors of scalars and arbitrary coupling constants.} Since the flavor-ordered partial amplitudes for multi-trace operators are not commonly encountered in the literature, in Appendix \ref{app:multi} we provide a definition of multi-trace partial amplitudes, which is relevant also at ${\cal O}(p^4)$.

\subsection{A Mixed Theory?}

\begin{figure}[t]
\centering
\includegraphics[width=10cm]{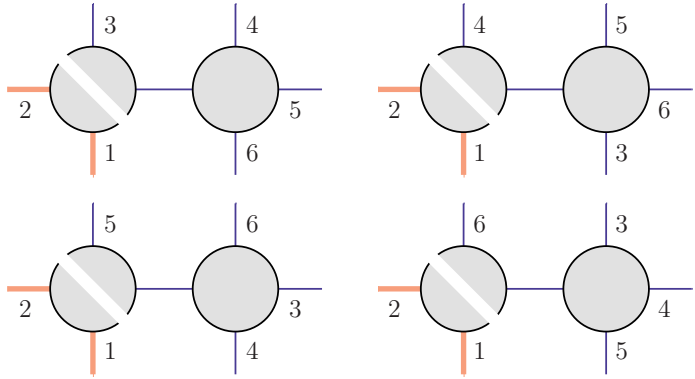}
\caption{\em The 4 factorization channels of $M (1,2|3,4,5,6)$.\label{fig:sb6sd}}
\end{figure}

Given that there are two different soft blocks at ${\cal O}(p^2)$, one could ask whether it is possible to construct an EFT using both soft blocks simultaneously. More specifically, using the single trace soft block in Eq. (\ref{eq:o2sos}) together with the double trace soft block in Eq. (\ref{eq:p2dtsb}), we could construct a 6-pt double-trace amplitude $M^{(2)} (1,2|3,4,5,6)$. There are  4 factorized blocks as shown in Fig. \ref{fig:sb6sd}, where we have also indicated the flavor flow. Using Eq.~(\ref{eq:sbrrl}) we have
\bea
M^{(2)} (1,2|3,4,5,6) &=& -\frac{ c_0\ d_0}{f^4} \ s_{12} \left[s_{46} \left(\frac{1}{P^2_{123}} + \frac{1}{P^2_{125}} \right)  +s_{35 } \left(\frac{1}{P^2_{124}} + \frac{1}{P^2_{126}} \right) \right] \non\\
&&+ M^{(2),\ctt} (1,2|3,4,5,6)\ ,
\eea
where the contact term is
\bea
&&M^{(2),\ctt} (1,2|3,4,5,6)\non\\
 &=& -\frac{ c_0\ d_0}{f^4}\sum_{i=1}^6 \Res_{z=1/a_i} \frac{1 }{z F_6(z) } \hats_{12} \left[\hats_{46} \left(\frac{1}{\hatP^2_{123}} + \frac{1}{\hatP^2_{125}} \right)  +\hats_{35 } \left(\frac{1}{\hatP^2_{124}} + \frac{1}{\hatP^2_{126}} \right) \right]\non\\
&=&-\frac{ c_0\ d_0}{f^4} \left\{ s_{46} \left[ \frac{a_3^3}{(a_3- a_1 ) (a_3 - a_2) (a_3 - a_5)} + \frac{a_5^3}{(a_5- a_1 ) (a_5 - a_2) (a_5 - a_3)} \right] \right.\non\\
&&\left.+ s_{35} \left[ \frac{a_4^3}{(a_4- a_1 ) (a_4 - a_2) (a_4 - a_6)} + \frac{a_6^3}{(a_6- a_1 ) (a_6 - a_2) (a_6 - a_4)} \right] + s_{12} \right\}.\label{eq:p2sd6c}
\eea
If the ``mixed'' EFT exists, this expression needs to be independent of the constants $A^{(r)}$ and $B$ when plugging in the general solution for $a_i$'s in Eq.~(\ref{eq:aigen}). 

It is easy to see that the contact interaction $M^{(2),\ctt} (1,2|3,4,5,6)$ is independent of $A^{(r)}$. However, it is {\em not} independent of $B$. This can be  verified using a set of  momenta numerically. For example, we can choose the momenta $p_i$ to be, in arbitrary units,
\bea
p_1 &=& (3, -2, 2, -1),\qquad p_2 = (-2, -2, 0, 0), \qquad p_3 = (3, 2, 2, 1),\non\\
p_4 &=& (2, -2, 0, 0),\qquad p_5 = \left(-\frac{1}3, 0, 0, \frac{1}3 \right), \qquad p_6 = \left(-\frac{17}3, 4, -4, -\frac{1}3 \right) \ ,
\eea
which satisfy $p_i^2=0$ and $\sum_i p_i=0$. The general solution for $a_i$, up to the overall scaling factor, can be written as
\bea
a_1 &=& B, \qquad a_2 = B+1, \qquad a_3 = B + \frac{4}5,\non\\
 a_4 &=& B+ \frac{3}5, \qquad a_5 = B- 2, \qquad a_6 = B+ \frac{2}5.
\eea
Plugging the above into Eq. (\ref{eq:p2sd6c}) we arrive at
\bea
M^{(2),\ctt} (1,2|3,4,5,6) = -\frac{c_0 d_0}{f^4} \left( \frac{32}9-\frac{100}3B-\frac{100}3B^2- \frac{275}{36}B^3 \right),
\eea
indicating that we cannot soft-bootstrap $M^{(2)} (1,2|3,4,5,6)$ using both soft blocks at ${\cal O}(p^2)$. Therefore, the single-trace and double-trace soft blocks at $\ordr (p^2)$ cannot co-exist and there is no consistent EFT that follows.

\section{Higher Orders in Derivative Expansion}
\label{sect:higherD}

So far we have seen that soft bootstrap allows one to construct two-derivative operators in EFT that are to all orders in $1/f$, with the only free parameter being $c_0$ or $d_0$, which can be absorbed into the overall normalization of the scale $f$. In the Lagrangian approach, these operators resum to a single nonlinear operator invariant under the transformation of the nonlinear shift symmetry \cite{Low:2014nga,Low:2014oga}. As is well-known, operators in the EFT of NLSM is organized
in terms of an increasing powers of derivatives. At the leading two-derivative order, there is only one operator whose coefficient  is fixed by the requirement of a canonically normalized scalar kinetic term. At higher orders in the derivative expansion, there exist several nonlinear operators in general, each with an incalculable Wilson coefficient encoding the unknown UV physics. Can the soft bootstrap program be extended to these higher derivative operators? How do the unknown Wilson coefficients emerge in the soft bootstrap? A particularly interesting class of operators is the Wess-Zumino-Witten (WZW) term that captures the effect of anomaly in a NLSM \cite{Wess:1971yu,Witten:1983tw}. Can the WZW term be soft-bootstrapped?

From the Lagrangian approach it seems the answer to above questions should be a definitive ``yes."  It is known that the Adler's zero condition corresponds to the Ward identity of the shift symmetry at the leading order in $1/f$, which in turn is associated with the existence of degenerate vacua and the phenomenon of spontaneous symmetry breaking \cite{Low:2018acv}.  In the following we study soft bootstrap at higher orders in the derivative expansion.

\subsection{General Remarks}
\label{subsect:general}

\begin{figure}[t]
\centering
\includegraphics[width=7.83cm]{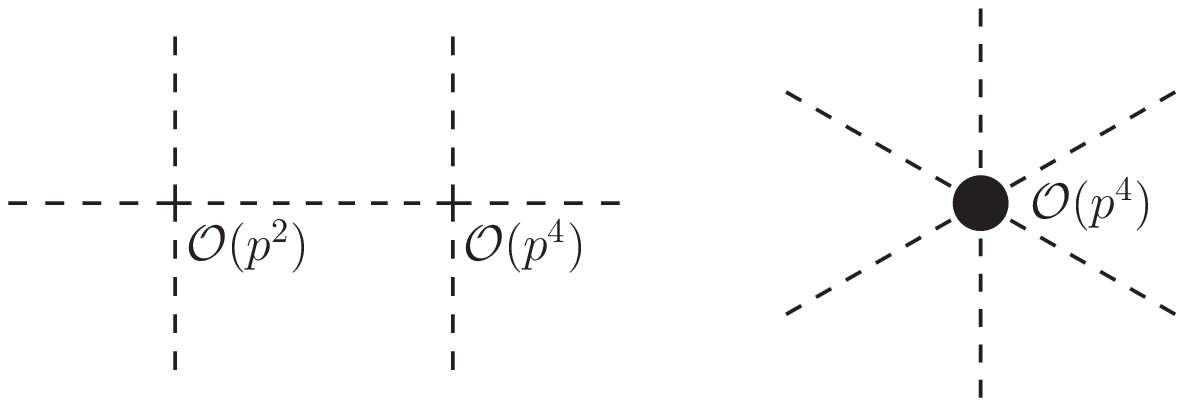}
\\
\includegraphics[width=12.73cm]{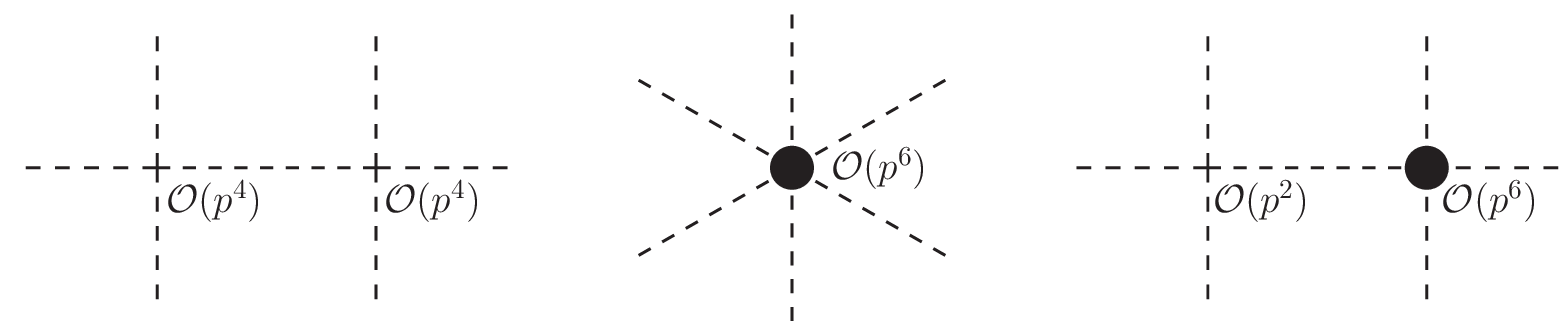}
\caption{\em Soft bootstrap in Feynman diagrams: black dots indicate unknown vertices that need to be fixed by the soft recursion. In the first row, only one unknown vertex enter the 6-pt amplitude at $\ordr (p^4)$. In the second row, two unknown vertices contribute to the 6-pt amplitude at $\ordr (p^6)$, whose coefficients remain undetermined after imposing the Adler's zero condition.\label{fig:op4va}}
\end{figure}

Before embarking on the pursuit of ${\cal O}(p^4)$ soft bootstrap,  we would like to understand what operators in Eq.~(\ref{eq:Lmultiscalar}) can be bootstrapped from an ${\cal O}(p^4)$ soft block? As an illustration, consider a 4-pt soft block containing four derivatives, ${\cal S}_4^{(4)}$, which has $n_i=d_i=4$ in Eqs.~(\ref{eq:nequationnew}) and (\ref{eq:dequationnew}). In terms of the derivative power counting parameter defined for single-scalar EFT's in Refs.~\cite{Cheung:2015ota,Cheung:2016drk,Elvang:2018dco}, it has
\be
\rho = \frac{d_i-2}{n_i-2} = 1 \ ,
\ee
which might suggest  other $\rho=1$ vertices can be soft-bootstrapped from the 4-pt soft block. At the 6-pt level, a $\rho=1$ vertex   contains 6 derivatives and the power counting based on $\rho$ would suggest its Wilson coefficient can be determined via soft bootstrap.  This intuition from single-scalar EFT's turns out to be incorrect. This is because there are two unknown vertices that would enter the 6-pt amplitude at the ${\cal O}(p^6)$: one has a multiplicity of 4 and the other has a multiplicity of 6, as shown in the second row of  Fig.~\ref{fig:op4va}. Therefore, imposing the Adler's zero condition cannot determine the individual Wilson coefficients of these two ${\cal O}(p^6)$ vertices. In fact, the essence of the soft bootstrap program relies on the property that, at a given order in the derivatives and the multiplicity of external particles, only one unknown vertex would enter the amplitude, which is demonstrated in the first row of Fig.~\ref{fig:op4va}. Consequently, the Adler's zero condition uniquely determines the coefficient of the unknown vertex. The program fails when two unknown vertices enter at the same time.

In fact, intuitions from the Lagrangian approach makes it clear that the nonlinear shift symmetry relates operators containing the same number of derivatives but an arbitrary number multiplicity in external fields \cite{Low:2014nga,Low:2014oga}.

To understand the situation more properly, one should go back to Eq.~(\ref{eq:dequationnew}), where one sees diagrams with a single insertion of ${\cal O}(p^4)$ vertex, with all other vertices being ${\cal O}(p^2)$, would carry $d=4$ and remain at ${\cal O}(p^4)$ regardless of the multiplicity of external particles, as shown in the 6-pt example given in Fig. \ref{fig:op4va}. In other words, by considering tree amplitudes containing a single insertion of lower-multiplicity ${\cal O}(p^4)$ vertex, with all other insertions at ${\cal O}(p^2)$, one could determine the ${\cal O}(p^4)$ vertex at a higher multiplicity by imposing the Adler's zero condition, in a fashion much similar to the soft bootstrap at ${\cal O}(p^2)$.  By repeating the reasoning iteratively, we expect that all operators in Eq.~(\ref{eq:Lmultiscalar}) of the form
\be
\label{eq:softopep4}
\frac{1}{\Lambda^2 f^{2k-2}} \partial^4 \left[ \Phi\right]^{2k}  
\ee
can be soft-bootstrapped from ${\cal S}_4^{(4)}$, consistent with the expectation from nonlinear shift symmetry. In a general NLSM, operators with different $\rho$ mix under soft bootstrap.

Using 4-pt and 5-pt soft blocks,  we can start building 6-pt and higher-pt amplitudes  using the soft recursion relation. Note that $\hatm_n (z)$ in the integrand of Eq. (\ref{eq:cauc}) is $\ordr (z^4)$ at large $z$, so that the integrand is at $\ordr (z^{3-n})$, and there is no pole at $z \to \infty$ for $n\ge 6$. Therefore, the on-shell recursion relation is still valid and can be used for soft bootstrap.

\subsection{4-pt Soft Blocks at ${\cal O}(p^4)$}
\label{sec:sbo4}

The 4-pt soft blocks at ${\cal O}(p^4)$ can again be classified as ``single-trace'' and ``double-trace'' soft blocks, and there are two independent soft blocks in each class:
\begin{align}
& &\text{Single-trace:}\ \ {\cal S}_1^{(4)} (1,2,3,4) &=  \frac{c_1}{\Lambda^2 f^2} \  s_{13}^2 \ , \ \  & {\cal S}_2^{(4)} (1,2,3,4)&=   \frac{c_2}{\Lambda^2 f^2} \  s_{12} s_{23}\ ,\label{eq:sa4s}\\
&&\text{Double-trace:}\ \ {\cal S}_1^{(4)} (1,2|3,4) &=   \frac{d_1}{\Lambda^2 f^2}  \ s_{12}^2 \ , \ \  & {\cal S}_2^{(4)} (1,2|3,4)&=    \frac{d_1}{\Lambda^2 f^2} \ s_{13} s_{23}\ ,\label{eq:sa4d}
\end{align}
where we have introduced four free parameters: $c_1, c_2, d_1$ and $d_2$. The power counting of mass scales in these soft blocks is given by Eq.~(\ref{eq:Lmultiscalar}). Notice that these soft blocks not only satisfy the Adler's zero condition, but the soft degrees of freedom seem to be enhanced due to the increasing power of momenta. However, at this order the soft blocks themselves are not on-shell amplitudes, which still only vanish linearly in the soft momentum.

In Section \ref{sec:op2so} we showed that there is no mixed theory at ${\cal O}(p^2)$: one either starts with the single trace soft block controlled by $c_0$ or the double trace soft block in $d_0$. Since the soft-bootstrap of ${\cal O}(p^4)$ vertices also involve ${\cal O}(p^2)$ vertices, as discussed in Section \ref{subsect:general}, we need to consider $c_0=0$ and $d_0=0$  separately.

\begin{figure}[t]
\centering
\includegraphics[width=9.7cm]{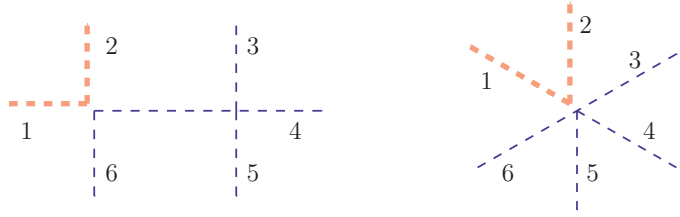}
\caption{\em Two classes of Feynman diagrams of $M (1,2|3,4,5,6)$. The diagram on the left contains a propagator, and there are 4 diagrams of this kind, corresponding to the 4 factorization channels. The diagram on the right is the contact term.\label{fig:fd6sd}}
\end{figure}

We start with the case $ c_0 \ne 0$ and $d_0 = 0$.  Using the soft blocks in Eqs. (\ref{eq:sa4s}) and (\ref{eq:sa4d}) we can construct 6-pt amplitudes with two different flavor orderings at ${\cal O}(p^4)$: $M (1,2,3,4,5,6)$ and $M(1,2|3,4,5,6)$, which are single trace and double-trace, respectively. Using Eq. (\ref{eq:sbrrl}), we calculate both 6-pt amplitudes analytically up to ${\cal O}(p^4)$:
\bea
&&M (1,2,3,4,5,6) \non\\
 &=& M^{(2)} (1,2,3,4,5,6) -\frac{c_0}{f^2} \left\{ \frac{c_1}{\Lambda^2f^2} \left[\frac{s_{13}  s_{46}(s_{13} + s_{46})}{P^2_{123}} + \frac{s_{24}  s_{15}(s_{24} + s_{15}) }{P^2_{234}}\right.\right. \non\\
&&\left.
+ \frac{s_{35} s_{26} ( s_{35} + s_{26} ) }{P^2_{345}} - \left( P^2_{135} \right)^2 - s_{13} s_{46} - s_{24} s_{15} - s_{35} s_{26}  \right]\non\\
&&+\frac{c_2}{\Lambda^2f^2}\left(\frac{s_{13}s_{45} s_{56} + s_{46} s_{12} s_{23}}{P^2_{123}} + \frac{s_{24}s_{56} s_{16} + s_{15} s_{23} s_{34}}{P^2_{234}} 
+ \frac{s_{35}s_{16} s_{12} + s_{26} s_{34} s_{45}}{P^2_{345}} \right.\non\\
&&\left.\left. \phantom{\frac{s_{13}}{P^2_{123}}}- P_{134}^2 s_{45} -s_{12} P^2_{146} - s_{12} s_{45}  - s_{14} s_{25} + s_{15} s_{24}  \right) \right\} + \ordr (p^6)\ ,
\eea
where $M^{(2)} (1,2,3,4,5,6)$ is the single trace 6-pt amplitude at ${\cal O}(p^2)$ in Eq.~(\ref{eq:single6pt}), and
\bea
&&M(1,2|3,4,5,6) \non\\
&=& -\frac{c_0}{f^2} \left\{\frac{d_1}{\Lambda^2f^2} \left[ s_{12}^2\left[ s_{46} \left(\frac{ 1}{P^2_{123}} + \frac{ 1}{P^2_{125}} \right) +s_{35} \left(\frac{ 1}{P^2_{124}} + \frac{ 1}{P^2_{126}}  \right) -1 \right] -s_{12} (s_{35} + s_{46})\right]  \right.\non\\
&&  +\frac{d_2}{\Lambda^2f^2} \left[ s_{46} \left(\frac{s_{13} s_{23}}{P^2_{123}} + \frac{s_{15} s_{25}}{P^2_{125}}  \right)  \left.+ s_{35} \left(\frac{s_{14} s_{24}}{P^2_{124}} + \frac{s_{16} s_{26}}{P^2_{126}}  \right) - (s_{15} + s_{13}) (s_{25} + s_{23}) + s_{12} s_{35}  \right] \right\}\non\\
&& + \ordr (p^6)\ .
\eea
The factorization channels, as well as the Feynman diagrams, of the above two kinds of flavor orderings are the same as in the $\ordr (p^2)$ case, which are shown in Figs. \ref{fig:6pt}, \ref{fig:st6pfc}, \ref{fig:sb6sd} and \ref{fig:fd6sd}. Notice that there is no double-trace ${\cal O}(p^2)$ amplitude because we started with $d_0=0$.

Going up to 8-pt amplitudes, there are three different flavor orderings: 
\be
M(1,2,3,4,5,6,7,8)\ , \quad M(1,2|3,4,5,6,7,8)\ , \quad M(1,2,3,4|5,6,7,8) \ ,
\ee
which can be built using Eq.~(\ref{eq:sbrr}). We checked numerically that the amplitudes are consistent and independent of $A^{(r)}$ and $B$ in the general solutions of $a_i$.  This indicates a consistent EFT can be built using the soft blocks
\be
\text{EFT}_1: \{c_0, c_1, c_2, d_1, d_2\}\ .
\ee 
At this order in derivative expansion, EFT$_1$ contains 4 free parameters $\{c_1, c_2, d_1, d_2\}$ in Eqs.~(\ref{eq:sa4s}) and (\ref{eq:sa4d}), with $c_0$ being absorbed into the normalization of $f$.   We will see in Section \ref{sec:mat} that these four free parameters correspond precisely to the four Wilson coefficients in the $\SU (N)$ NLSM at ${\cal O}(p^4)$ order.

Next we consider the other case: $d_0 \ne 0$ and $c_0 = 0$. There are two flavor orderings at ${\cal O}(p^4)$: $M(1,2|3,4,5,6)$ and $M (1,2|3,4|5,6)$ which correspond to double-trace and triple-trace amplitudes, respectively. Using Eq.~(\ref{eq:sbrrl}) again we find the double-trace amplitude $M(1,2|3,4,5,6)$ bootstrapped from $c_1$ and $c_2$ is not consistent. On the other hand, the triple-trace amplitude built from $d_1$ and $d_2$  does exist:
\bea
&&M (1,2|3,4|5,6)\non\\
&=& M^{(2)} (1,2|3,4|5,6)- \frac{d_0}{f^2} \left\{ \frac{d_1}{\Lambda^2 f^2} \left[ s_{12} s_{56}(s_{12} + s_{56}) \left( \frac{1}{P^2_{123}} + \frac{1}{P^2_{124}} \right) \right. \right.\non\\
&&+ s_{34} s_{56}(s_{34} + s_{56}) \left( \frac{1}{P^2_{134}} + \frac{1}{P^2_{234}} \right)\non\\
&& \left. + s_{12} s_{34}(s_{12} + s_{34}) \left( \frac{1}{P^2_{125}} + \frac{1}{P^2_{126}}  \right) -(s_{12} + s_{34} + s_{56})^2\right]\non\\
&&+\frac{d_2}{\Lambda^2 f^2} \left( \frac{s_{13} s_{23} s_{56} + s_{12} s_{45} s_{46}}{P^2_{123}} + \frac{s_{14} s_{24} s_{56} + s_{12} s_{35} s_{36}}{P^2_{124}}+\frac{s_{13} s_{14} s_{56} + s_{25} s_{26} s_{34}}{P^2_{134}}\right.\non\\
&&\phantom{\frac{s_2}{P^2_6}} +\frac{s_{23} s_{24} s_{56} + s_{15} s_{16} s_{34}}{P^2_{234}}   +\frac{s_{15} s_{25} s_{34} + s_{36} s_{46} s_{12}}{P^2_{125}}+\frac{s_{16} s_{26} s_{34} + s_{35} s_{45} s_{12}}{P^2_{126}}-P^2_{123} P^2_{124} \non\\
&&\left. \left.\phantom{\frac{s_2}{P^2_6}}- P^2_{125} P^2_{126} - P^2_{156} P^2_{256} +s_{12} s_{56} + s_{34} s_{56} + s_{12} s_{34} \right)\right\} + \ordr (p^6).\qquad
\eea
Again, the factorization channels and Feynman diagrams are  identical to those in the $\ordr (p^2)$ case, which are shown in Figs. \ref{fig:sb6d} and \ref{fig:fd6dou}. The 8-pt amplitude $M^{(4)} (1,2|3,4|5,6|7,8)$ built using Eq. (\ref{eq:sbrr}) is of quadruple-trace and also exists. Going to $2n$-pt amplitude, it always contains $n$ traces in the flavor ordering. In the end we arrive at a second consistent EFT in soft bootstrap,
\be
\text{EFT}_2: \{ d_0, d_1, d_2\} \ ,
\ee
and it has two free parameters $\{d_1, d_2\}$ in Eq. (\ref{eq:sa4d}). In Section \ref{sec:mat} we will match EFT$_2$ to the $\SO (N)$ NLSM, which has two Wilson coefficients at ${\cal O}(p^4)$.

\subsection{5-pt Soft Blocks: Wess-Zumino-Witten Terms}
\label{sec:wzw}

In this section we consider soft blocks with $5$ external legs at ${\cal O}(p^4)$. We find one single trace soft block that is parity-odd,
\be
{\cal S}_-^{(4)} (1,2,3,4,5) = \frac{c_-}{\Lambda^2 f^3}\  \vep (1234)\ , \qquad \vep (ijkl) \equiv \vep_{\mu \nu \rho \sigma} p_i^\mu p_j^\nu p_k^\rho p_l^\sigma\ . \label{eq:sa5}
\ee 
The expression is invariant under cyclic permutations upon total momentum conservation.  ${\cal S}_-^{(4)}$  clearly corresponds to the Wess-Zumino-Witten (WZW) term \cite{Wess:1971yu,Witten:1983tw}, which accounts for the anomaly that may arise in a NLSM in $D=4$.  It is well-known that in the CCWZ construction the existence of WZW term, or the lack thereof, depends on the existence of a rank-5 totally anti-symmetric invariant tensor in the coset $G/H$ \cite{DHoker:1994rdl,DHoker:1995mfi}. Such information is clearly not available in soft bootstrap. Nevertheless we will see soon that the group-theoretic considerations based on $G/H$ can be exactly reproduced in a remarkable way, after taking into account the Bose symmetry in the IR.

\begin{figure}[t]
\centering
\includegraphics[width=7.83cm]{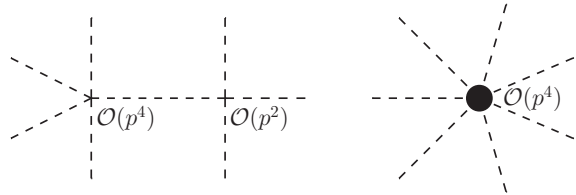}
\caption{\em Soft bootstrap for the 7-pt vertex in the WZW term. There are two classes of diagrams and the Adler's zero condition fixes the 7-pt contact interaction.\label{fig:sbwzwp4}}
\end{figure}

The WZW term has been considered previously in Refs. \cite{Cheung:2016drk,Elvang:2018dco}, however,  only  the leading 5-pt vertex in $1/f$ expansion was discussed.  Here we are interested in soft-bootstrapping higher-pt amplitudes that are of $\ordr(p^4)$, using the WZW soft block. These amplitudes correspond to interactions that are higher orders in $1/f$ in the WZW term. What vertices can be soft-bootstrapped from ${\cal S}_-^{(4)} (1,2,3,4,5)$? Again from Eq.~(\ref{eq:dequationnew}) one can see that all diagrams with one insertion of ${\cal S}_-^{(4)} (1,2,3,4,5)$ and an arbitrary number of two-derivative vertices will carry the same number of derivatives as ${\cal S}_-^{(4)} (1,2,3,4,5)$. As a consequence, soft bootstrap can be used to constrain  operators of the form
\be
\frac{1}{\Lambda^2 f^{3+2k}}[\Phi]^{2k}\, \vep_{\mu\nu\rho\sigma }\partial^\mu\partial^\nu\partial^\rho\partial^\sigma [\Phi]^5 \ .
\ee
For example, there is only one unknown vertex, the 7-pt contact interaction, contained in the Feynman diagrams contributing to the 7-pt amplitude, as shown in Fig. \ref{fig:sbwzwp4}. The Adler's zero condition then fixes the 7-pt vertex uniquely. These operators make up the  WZW term to all orders in $1/f$.

Again, we need to discuss separately EFT$_1$ and EFT$_2$. In EFT$_1$ we use ${\cal S}_-^{(4)} (1,2,3,4,5)$ together with ${\cal S}^{(2)}(1,2,3,4)$ in Eq.~(\ref{eq:o2sos}) to construct  higher-pt single-trace amplitudes, which give consistent higher-pt amplitudes. For example the 7-pt amplitude is calculated analytically using Eq. (\ref{eq:sbrrl}):
\bea
&&M(1,2,3,4,5,6,7)\non\\
&=& -\frac{c_0}{f^2} \frac{c_-}{\Lambda^2 f^3} \left[ \frac{\vep (1234) s_{57}}{P^2_{567}} + \frac{\vep (2345) s_{16}}{P^2_{671}} + \frac{\vep (3456) s_{27}}{P^2_{127}} + \frac{\vep (4567) s_{13}}{P^2_{123}} + \frac{\vep (5671) s_{24}}{P^2_{234}} \right. \non\\
&&\left. + \frac{\vep (6712) s_{35}}{P^2_{345}}+ \frac{\vep (7123) s_{46}}{P^2_{456}} - \vep (3456) - \vep (1456) - \vep (1256) - \vep (1236) - \vep (1234)\right] \non\\
&& + \ordr (p^6)\ .\label{eq:a7psu}
\eea
At the 9-pt amplitude we have verified that the amplitude built recursively is independent of the arbitrary constants $A^{(r)}$ and $B$ in the general solution of $a_i$'s. This suggests  we can soft-bootstrap all $(5+2n)$-pt amplitudes of the full WZW term of $\SU (N)$ NLSM. On the other hand, in EFT$_2$ the recursively constructed 7-pt amplitude $M(1,2, 3,4,5 | 6,7)$ is not a consistent amplitude, in general. More explicitly, the 7-pt amplitude is given by
\bea
M (1,2,3,4,5|6,7) &=& -\frac{  d_0}{f^2} \frac{c_-}{\Lambda^2 f^3} \ s_{67} \left[ \frac{\vep (1234) }{P^2_{567}} + \frac{\vep (2345) }{P^2_{167}} + \frac{\vep (3451) }{P^2_{267}} + \frac{\vep (4512) }{P^2_{367}} + \frac{\vep (5123) }{P^2_{467}}\right] \non\\
&&+ M^{(4),\ctt} (1,2,3,4,5|6,7) + \ordr (p^6),
\eea
where the contact term is
\bea
&&M^{(4),\ctt} (1,2,3,4,5|6,7)\non\\
&=& -\frac{  d_0}{f^2} \frac{c_-}{\Lambda^2 f^3} \sum_{i=1}^7 \Res_{z=1/a_i} \frac{1 }{z F_7(z) } \hats_{67} \left[ \frac{\hat{\vep} (1234) }{\hatP^2_{567}} + \frac{\hat{\vep} (2345) }{\hatP^2_{167}} + \frac{ \hat{\vep} (3451) }{\hatP^2_{267}} \right.\non\\
&&\left. + \frac{\hat{\vep} (4512) }{\hatP^2_{367}} + \frac{\hat{\vep} (5123) }{\hatP^2_{467}}\right]  \non\\
&=&\frac{  d_0}{f^2} \frac{c_-}{\Lambda^2 f^3} \left\{  \frac{a_5^2 \vep (1234)}{(a_5- a_6 ) (a_5 - a_7)} + \frac{a_1^2 \vep (2345)}{(a_1- a_6 ) (a_1 - a_7)} +\frac{a_2^2 \vep (3451)}{(a_2- a_6 ) (a_2 - a_7)}  \right.\non\\
&&\left.+\frac{a_3^2 \vep (4512)}{(a_3- a_6 ) (a_3 - a_7)} +\frac{a_4^2 \vep (5123)}{(a_4- a_6 ) (a_4 - a_7)}  \right\}.\label{eq:7ptdtc}
\eea
It is easy to check numerically that the $a_i$ dependence does not cancel in the above when plugging in the general solution in Eq.~(\ref{eq:aigen}). This indicates the absence of the WZW term in $\SO (N)$ NLSM, in general.

There is  a subtlety in the preceding arguments, which involves the number of flavors  $N_f$ and the Bose symmetry. If  $N_f < 5$ in the EFT,  two or more scalars in Eq.~(\ref{eq:sa5}) are identical and  Bose symmetry requires the amplitude must be symmetric in external momenta of identical scalars. As a result, the WZW soft block vanishes due to the anti-symmetric Levi-Civita tensor used in contracting the external momenta. Therefore, we arrive at the important observation:
\begin{itemize}
\item ${\cal S}_-^{(4)}$ is non-vanishing only if the number of flavors $N_f \ge 5$.
\end{itemize}
For the $\SU (N)$ NLSM the number of flavors is $N^2-1$, which implies the WZW term exists only for $N\ge 3$.

The interesting interplay between $N_f$ and the Bose symmetry continues at higher-pt, as $n$-pt amplitudes with $n>N_f$ always contain identical scalars. The amplitude must then be symmetric in arbitrary permutations of external momenta of the identical scalars, in addition to the cyclic ordering imposed by the partial amplitudes. Such a requirement might render an otherwise inconsistent amplitude consistent. A case in point is  the 7-pt WZW amplitude in EFT$_2$, which was shown to be inconsistent in general. However,  when $N_f=5$,  $M(1,2,3,4,5|6,7)$ now contains at least three scalars of identical flavors.\footnote{$N_f = 6$, on the other hand, can contain only two scalars of identical flavors. Therefore, if we choose $\{6,7\}$ to be identical scalars, as $\{6,7\}$ is already symmetrized, the resulting amplitude is no different from a generic amplitude of $N_f \ge 7$.} Without loss of generality, let us assume particles $\{ 5,6,7 \}$ have the same flavor. Then the 7-pt amplitude satisfying both the cyclic ordering and the Bose symmetry is
\be
M (1,2,3,4,\{ 5,6,7 \} ) = M(1,2, 3,4,5 | 6,7) + M(1,2, 3,4,6 | 5,7) + M(1,2, 3,4,7 | 5,6)\ ,
\ee
where $\{\cdots\}$ in the left-hand side (LHS) of the above denotes a group of external states with identical flavor.\footnote{There are no 7-pt amplitudes of other flavor structures, like $M (1,2,3,\{4,5\},\{6,7\})$: no factorization channels exist, and no contact terms at $\ordr (p^4)$ that satisfy Adler's zero condition exist.} Under such a symmetrization of $\{5,6,7 \}$,
\be
M^{(4),\ctt} (1,2,3,4,\{ 5,6,7 \} ) = -\frac{  d_0}{f^2} \frac{c_-}{\Lambda^2 f^3}  \sum_{i=1}^7 \Res_{z=1/a_i} \frac{1 }{z F_7(z) }  \hat{\vep} (1234) = \frac{  d_0}{f^2} \frac{c_-}{\Lambda^2 f^3} \vep (1234)\ ,
\ee
so that
\bea
\label{eq:so57pa}
M (1,2,3,4,\{ 5,6,7 \} )  &=& -\frac{d_0}{f^2} \frac{c_-}{\Lambda^2 f^3} s_{67} \left[ \frac{\vep (2345) }{P^2_{167}} + \frac{\vep (3451) }{P^2_{267}}+ \frac{\vep (4512) }{P^2_{367}} + \frac{\vep (5123) }{P^2_{467}}\right] \non\\
&& + ( 5 \leftrightarrow 6) + ( 5 \leftrightarrow 7) + \ordr (p^6)\ .
\eea
We see, remarkably, the $a_i$ dependence in Eq. (\ref{eq:7ptdtc}) is canceled out and a consistent 7-pt amplitude now exists!

\begin{figure}[t]
\centering
\includegraphics[width=13.73cm]{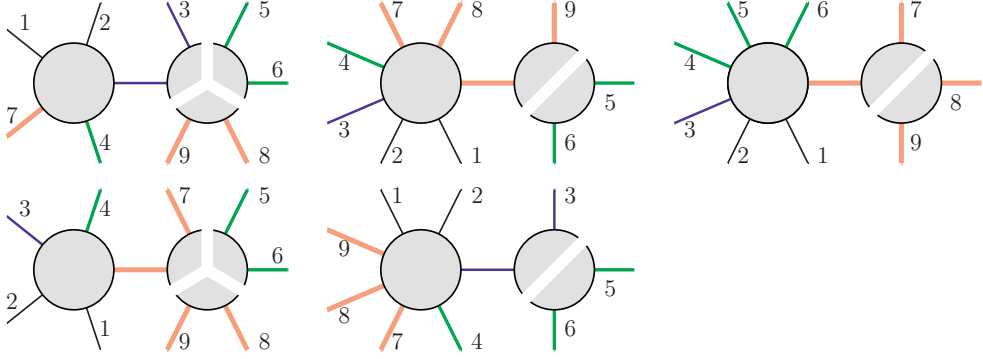}
\caption{\em Five classes of factorized blocks in $M \left(1,2,3,\{ 4,5,6 \}, \{7,8,9\} \right)$. Except for the black, thin lines, the legs of the same color in the above have the same flavor in the amplitude. We need to sum over different permutations of the above channels to get the correct ordering property. For example, the final amplitude needs to be totally symmetric in $\{4,5,6\}$, thus we need to sum up the first type of factorization channels in the above where the green line attached to the 5-pt soft block are $4$, $5$ or $6$. There are a total of 71 distinct channels involved.\label{fig:so59p3}}
\end{figure}

\begin{figure}[t]
\centering
\includegraphics[width=9.13cm]{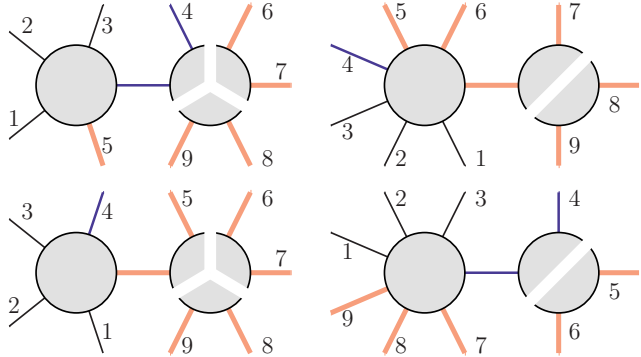}
\caption{\em Four classes of factorized blocks in $M \left(1,2,3,4, \{5,6,7,8,9\} \right)$. 71 distinct factorization channels contribute to this amplitude. See also the caption in Fig.~\ref{fig:so59p3}. \label{fig:so59p5}}
\end{figure}

At the $9$-pt amplitudes, there are two possible flavor structures, 
\be
M \left(1,2,3,\{ 4,5,6 \}, \{7,8,9\} \right) \ , \qquad M \left(1,2,3,4, \{5,6,7,8,9\} \right) \ . 
\ee
To construct them recursively we need the following amplitudes
\be
M(1,2|3,4)\ , \quad M(1,2|3,4|5,6)\ , \quad M(1,2,3,4,5)\ , \quad M \left(1,2,3,4,\{ 5,6,7 \} \right)\ .
\ee 
The factorization channels for $M \left(1,2,3,\{ 4,5,6 \}, \{7,8,9\} \right)$ and $M \left(1,2,3,4, \{5,6,7,8,9\} \right)$ are shown in Figs. \ref{fig:so59p3} and \ref{fig:so59p5}, respectively. We have checked numerically that both of the 9-pt amplitudes can be constructed consistently using Eq. (\ref{eq:sbrr}), leading to the conclusion that there is a WZW term for the $\SO (5)$ NLSM.

One clarifying remark regarding the flavor structure of the WZW amplitudes in EFT$_2$ is warranted. There is only one WZW soft block ${\cal S}^{(4)}_-(1,2,3,4,5)$, which is of ``single trace'' and therefore invariant under the cyclic permutation of external legs. In $M (1,2,3,4,\{ 5,6,7 \} )$, since the flavors in $\{5,6,7\}$ are identical, the amplitude is also invariant under the cyclic permutation of $(1234j)$ for $j\in \{5,6,7\}$. Similarly, the 9-pt amplitude $M \left(1,2,3,\{ 4,5,6 \}, \{7,8,9\} \right)$ is invariant under the cyclic permutation of $(123jk)$ for $j\in\{4,5,6\}$ and $k\in\{7,8,9\}$. The same comment also applies to $M \left(1,2,3,4, \{5,6,7,8,9\} \right)$. In the end the WZW amplitude in EFT$_2$ for $N_f=5$ has only one flavor structure that involves the cyclic permutation of the 5 distinct flavors.

\section{Matching to Lagrangians}
\label{sec:mat}

Having constructed  EFT$_1$ and EFT$_2$ in soft-bootstrap up to ${\cal O}(p^4)$, we match these two theories to effective Lagrangians of NLSM in this section. The general and top-down approach in the Lagrangian formulation for such effective interactions is given by Callan, Coleman, Wess and Zumino half-a-century ago \cite{Coleman:1969sm,Callan:1969sn}. The CCWZ construction requires knowledge of a spontaneously broken group $G$ in the UV and an unbroken group $H$ in the IR. The generators of $G$ include the ``unbroken generators'' $T^i$, which are associated with $H$, as well as the ``broken generators'' $X^{a}$, which are associated with the coset $G/H$. The NGB's are then coordinates parameterizing the coset $G/H$. We will sometimes refer to CCWZ as the ``the coset construction.''

At first sight it may seem rather improbable that soft bootstrap could (re)construct effective Lagrangians in the CCWZ approach, since one makes no reference to a spontaneously broken group $G$ in the UV in soft bootstrap; all that is needed is the Adler's zero condition, an IR property of  on-shell amplitudes. Indeed, the coset construction completely obscures the ``infrared universality:''  effective interactions of NGB's are dictated by their quantum numbers in the IR and independent of the broken group $G$ in the UV.

Only recently was it realized that an IR construction of effective Lagrangians exists, without reference to the spontaneously broken symmetry $G$, which makes use of nonlinear ``shift symmetries'' acting on a set of  massless scalars $\pi^a$ furnishing a linear representation of the (unbroken) group $H$ \cite{Low:2014nga,Low:2014oga}. It turns out that imposing the shift symmetry in the Lagrangian is equivalent to imposing Adler's zero condition on the on-shell amplitudes, which arises as a consequence of the Ward identity for  the shift symmetry \cite{Low:2015ogb,Low:2017mlh}. In this sense the IR construction  can be viewed as the realization of soft bootstrap in the Lagrangian formulation \cite{Low:2018acv}. Since the IR construction is more similar to the soft bootstrap program in philosophy,  we will adopt the IR approach to consider effective Lagrangians corresponding to EFT$_1$ and EFT$_2$.

\subsection{The leading two-derivative Lagrangian}
\label{app:infc}

As a warm-up exercise to the eventual discussion of ${\cal O}(p^4)$ operators, as well as to set the notation, we briefly consider the leading two-derivative Lagrangian of NLSM. Consider a set of scalars $\pi^a$ transforming as a linear representation of an unbroken group $H$. Introducing the bra-ket notation $(|\pi\rangle)_a=\pi^a$, we have
\be
|\pi\rangle \to e^{i\alpha_i T^i} |\pi\rangle \ ,
\ee
where $T^i$ is the generator of $H$ in the particular representation under consideration. Moreover, we will choose a basis such that $T^i$ is purely imaginary and anti-symmetric:
\be
(T^i)_{ab} = -(T^i)_{ba} \ , \qquad (T^i)^* = - T^i \ .
\ee
We are interested in constructing an effective Lagrangian invariant under the following nonlinear shift symmetry \cite{Low:2014nga,Low:2014oga},
\be
\label{eq:shiftall}
|\pi \> \to |\pi\> + \sqrt{\mt} \cot \sqrt{\mt}\  |\vep \>\ , \qquad
(\mt)_{ a  b} = \frac{1}{f^2} (T^i)_{ a c} (T^i)_{d b}\  \pi^c \pi^{d} \ ,
\ee
where $(|\vep \>)_a = \vep^a$ represents an infinitesimal constant ``shift'' in $\pi^a$. Eq.~(\ref{eq:shiftall}) at the leading order is simply
\be
\pi^a \to \pi^a +\vep^a + {\cal O}\left(\frac1{f^2}\right) \ ,
\ee
whose Ward identity leads to the Adler's zero condition \cite{Low:2015ogb,Low:2017mlh}. Terms that are higher order in $1/f$ are dictated by the unbroken $H$-symmetry and the vanishing of $n$-pt tree amplitudes among identical massless scalars in Eq.~(\ref{eq:Msinglef}).

The building block of effective Lagrangians consists of two objects:
\bea
|d_\mu \> &=&  \frac{1}{f} F_1 (\mt)| \partial_\mu \pi \>\ ,\label{eq:irdefd}\\
E_\mu^i &=&  \frac{1}{f^2} \< \partial_\mu \pi| F_2 (\mt)| T^i \pi \>\ ,\label{eq:irdefe}
\eea
where
\be
F_1 (\mt) = \frac{\sin \sqrt{\mt}}{\sqrt{ \mt} } \ , \qquad F_2 (\mt ) = -\frac{2i}{\mt} \sin^2 \frac{\sqrt{\mt}}{2} \  .
\ee
In the above $d_\mu^a$ transforms covariantly under the shift symmetry in Eq.~(\ref{eq:shiftall}) while $E_\mu^i$ transforms in the adjoint representation of $H$ like a ``gauge field,''
\bea
\label{eq:dmutransf}
|d_\mu \> &\to& h_A (\vep, \pi)\ |  d_\mu\> ,\\
E_\mu^i T^i &\to&\  h_A (\vep, \pi)\  E_\mu^i T^i\ h_A^\dagger (\vep, \pi) - i\, h_A (\vep, \pi)\ \partial_\mu h_A^\dagger (\vep, \pi) ,\label{eq:eitransf}
\eea
where the specific form of $h_A (\vep, \pi)$ does not concern us here. Then the leading two-derivative operator is unique:
\bea
\lag^{(2)} = \frac{f^2}{2} \< d_\mu d^{\mu} \>\ , \label{eq:nlsmir}
\eea
where the coefficient is fixed by canonical normalization of the scalar kinetic term.

A general discussion on four-derivative operators in the NLSM effective action is delegated to Appendix \ref{app:allop}. We note here  that in the literature they have been enumerated in two contexts: chiral Lagrangian in low-energy QCD \cite{Gasser:1983yg, Gasser:1984gg} and nonlinear Lagrangian for a composite Higgs boson \cite{Contino:2011np,Panico:2015jxa,Liu:2018vel,Liu:2018qtb}. In the former case $\pi^a$'s furnish the adjoint representation of $\SU (N)$ group, while in the latter $\pi^a$'s transform as the fundamental representation of $\SO (N)$ group. It turns out that these are precisely the two cases that are matched to EFT$_1$ and EFT$_2$, respectively, which we will consider in the next two subsections.

We would like to finish this subsection with the power counting scheme based on the naive dimensional analysis (NDA) \cite{Manohar:1983md}. In the EFT of NLSM, each derivative $\partial_\mu$, and as a result, $d_\mu^a$ and $\nabla_\mu = \partial_\mu + iE_\mu^i T^i$, is suppressed by an energy scale $\Lambda$; each field $\pi^a$ is suppressed by the coupling constant $f$. The Lagrangian with a canonically normalized kinetic term is given by
\bea
\label{eq:powercount}
\lag =f^2 \Lambda^2\ \tilde{\lag} (d/\Lambda, \nabla/\Lambda)\ .
\eea
Requiring that the change in the coupling of a particular operator due to loop-induced effects to be comparable to the natural size dictated by power counting in Eq.~(\ref{eq:powercount}), one arrives at
\be
\Lambda \approx 4\pi f \ ,
\ee
which is the cutoff of the effective Lagrangian.

\subsection{Adjoint of $\SU (N)$}
\label{sec:sunl}
In this subsection we consider a set of massless scalars $\pi^a$'s which transform as the adjoint representation of $\SU (N)$ group.   In the CCWZ construction this scenario could arise from the coset $\SU (N)\times \SU(N)/\SU (N)$ and the $\ordr (p^2)$ tree amplitudes have been studied extensively from the on-shell perspective in Ref.~\cite{Kampf:2013vha}. As emphasized in Section \ref{sec:sbo}, the full amplitudes in this case have the nice property that the flavor factor factorizes simultaneously with the partial amplitudes defined in Eq.~(\ref{eq:stfod}), which can be seen as a consequence of the U(1) decoupling relations: if the coset is enhanced to  $\U (N)\times \U (N)$, vertices containing one U(1) NGB vanish. At $\ordr (p^4)$ the same arguments continue to hold and the partial amplitudes also factorize simultaneously with the flavor factor.

Given that $\pi^a$'s transform as the adjoint representation, we can write 
\be
\label{eq:twotrace}
d_\mu = d_\mu^a\, T^a \ , \qquad  {\cal L}^{(2)} = \frac{f^2}2 {\rm tr} (d_\mu d^\mu) \ ,
\ee
where $T^a$ is the generator of $\SU (N)$. We see that the leading two-derivative Lagrangian can be written as a single trace operator and the resulting partial amplitudes are symmetric in cyclic ordering of external particles. The two-derivative single trace soft block, ${\cal S}_4^{(2)}(1234)$, is precisely the 4-pt vertex following from Eq.~(\ref{eq:twotrace}) \cite{Kampf:2013vha} and the $\ordr(p^2)$ amplitudes bootstrapped from ${\cal S}_4^{(2)}(1234)$ are the corresponding $n$-pt partial amplitudes.

At $\ordr (p^4)$, as shown in Appendix \ref{app:allop}, there are  four ``parity-even'' operators for $\SU (N)$ in general:
\bea
O_1 &=& \left[\tr ( d_\mu d^\mu ) \right]^2\ ,\\
O_2 &=& \left[\tr ( d_\mu d_\nu ) \right]^2\ ,\\
O_3 &=& \tr ( [ d_\mu, d_\nu]^2 )\ ,\\
O_4 &=& \tr ( \{ d_\mu, d_\nu \}^2 )\ .
\eea
Notice that there are two double-trace operators $\{ O_1, O_2\}$ and two single-trace operators $\{ O_3, O_4\}$.  In addition, there is the ``parity-odd"  WZW term, which can be expressed using the action 
\bea
S_{\wzw} \propto \int d^5 y\ \vep^{\mu \nu \alpha \beta \gamma}\, \tr  ( d_\mu d_\nu d_\alpha d_\beta d_\gamma ) = \int d^4 x\ O_\wzw\ .
\eea
The WZW term for $\SU (N)$ is also a single trace operator. Then the $\ordr (p^4)$ Lagrangian   can be written as
\bea
\lag^{(4)} = \frac{f^2}{\Lambda^2} \left( \sum_{i=1}^4 C_i O_i + C_5 O_\wzw \right) ,\label{eq:sunp4l}
\eea
where $C_i$, $i=1,2,\cdots,5 $ are the unknown Wilson coefficients encoding the incalculable UV physics.

It is worth noting that, for $\SU (2)$  only two  out of the four parity-even operators are independent. This is easily seen by using properties of Pauli matrices in the adjoint of $\SU (2)$. For $\SU (3)$, three out of the four are independent. This can again be checked explicitly using the Gell-Mann matrices for $\SU (3)$.\footnote{A rigorous proof for the linear dependence of $O_4$ on $O_1$, $O_2$ and $O_3$ relies on the fact that there are no rank-4 totally symmetric invariant tensors in the adjoint indices of $\SU (2)$ or $\SU (3)$, except for ones constructed using Kronecker deltas \cite{deAzcarraga:1997ya,vanRitbergen:1998pn}.}

Recall that in EFT$_1$ there exist five free parameters from the five $\ordr (p^4)$ soft blocks: $c_1, c_2, d_1, d_2$ and $c_-$. We can match the partial amplitudes from Eq.~(\ref{eq:sunp4l}) with those from EFT$_1$. This is achieved by calculating the 4-pt interactions in the Lagrangian:
\bea
\lag^{(4)}&=&\frac{1}{f^2 \Lambda^2} \left\{ C_1 \left[ \tr \left( \partial_\mu \Pi \partial^\mu \Pi \right) \right]^2 + C_2 \left[ \tr \left( \partial_\mu \Pi \partial_\nu \Pi \right) \right]^2 \right.\non\\
&+&\left. 2(C_3 + C_4) \tr [ \left( \partial_\mu \Pi \partial_\nu \Pi \right)^2 ] + 2(C_4 - C_3) \tr [ \left( \partial_\mu \Pi \partial^\mu \Pi \right)^2 ] \right\} + \ordr \left(\frac{1}{f^4}\right)\ ,
\eea
where we have adopted the shorthand notation $\Pi \equiv \pi^a T^a$. Thus the 4-pt vertices are
\bea
V^{(4)} (1,2,3,4) &=& \frac{4}{f^2 \Lambda^2} \left[ (C_4 - C_3) \left(p_1 \cdot p_2\ p_3 \cdot p_4 + p_1 \cdot p_4 \ p_2 \cdot p_3 \right)\right.\non\\
&&\left. +2 (C_3 + C_4)p_1 \cdot p_3\ p_2 \cdot p_4 \right] ,\\
V^{(4)} (1,2|3,4) &=& \frac{4}{f^2 \Lambda^2} \left[ 2C_1 p_1 \cdot p_2 \ p_3 \cdot p_4  + C_2 ( p_1 \cdot p_3 \ p_2 \cdot p_4+ p_1 \cdot p_4 \ p_2 \cdot p_3 ) \right]  ,
\eea
generating the soft blocks
\bea
{\cal S}^{(4)}(1,2,3,4) &=& \frac{1}{f^2 \Lambda^2} \left[ (C_3+ 3C_4 ) s_{13}^2 + 2 (C_3 - C_4)s_{12} s_{23} \right],\non\\
{\cal S}^{(4)}(1,2|3,4)&=&\frac{1}{f^2 \Lambda^2} \left[ (2 C_1 + C_2) s_{12}^2 - 2C_2 s_{13} s_{23} \right].
\eea
Comparing with Eqs. (\ref{eq:sa4s}) and (\ref{eq:sa4d}), we see that
\bea
 c_1 =C_3+ 3 C_4 ,\qquad c_2 = 2(C_3- C_4),\qquad d_1 &=& 2C_1 + C_2,\qquad d_2 = -2C_2.
\eea
The two single-trace soft blocks, $c_1$ and $c_2$, could soft-bootstrap the amplitudes arising from the two single-trace operators $O_1$ and $O_2$, and similarly for the two double-trace soft blocks and double-trace operators.

Similarly, by calculating the 5-pt vertex contributed by the WZW term we get
\bea
V^{(4)} (1,2,3,4,5) = {\cal S}^{(4)}_- (1,2,3,4,5) = \frac{5 C_5}{f^3 \Lambda^2 }\, \vep (1234),
\eea
so that
\bea
  c_- &= &5C_5 .
\eea
The 7-pt local operator in the Lagrangian is
\bea
\frac{5 C_- }{21f^5 \Lambda^2} \vep^{\mu \nu \rho \sigma}  \tr \left( \Pi^3 \partial_\mu \Pi \partial_\nu \Pi \partial_\rho \Pi \partial_\sigma \Pi - 6 \Pi \partial_\mu \Pi \Pi^2 \partial_\nu \Pi \partial_\rho \Pi \partial_\sigma \Pi  -3\Pi^2 \partial_\mu \Pi \partial_\nu \Pi \Pi \partial_\rho \Pi \partial_\sigma \Pi \right),\non\\
\eea
which leads to the 7-pt vertex
\bea
V (1,2,3,4,5,6,7) = \frac{5C_-}{21f^3 \Lambda^2 } \left[ \vep (1234) + 6 \vep (1235) - 3 \vep (1245)  + \cycl \right],
\eea
with $\cycl$ denoting the terms generated by cyclic permutation of momentum indices $\{ 1,2, \cdots , 7\}$. The 7-pt amplitude calculated using the Feynman rules completely agrees with Eq. (\ref{eq:a7psu}).
We have seen in Section \ref{sec:wzw} that the corresponding soft block ${\cal S}_-^{(4)} (1,2,3,4,5)$ is non-zero only if $N_f\ge 5$. For the adjoint of $\SU (N)$, the number of flavor $N_f = N^2-1$, which implies the WZW term exists only for $N\ge 3$ \cite{DHoker:1994rdl,DHoker:1995mfi}.

In the end, we conclude
\be
\text{EFT$_1$  = $\SU (N)$ Adjoint NLSM}\ . \non
\ee
Moreover, the number of independent operators in the derivative expansion coincides with the number of independent soft blocks. On the other hand, coefficients in the $1/f$ expansion are completely determined by soft-bootstrap.

\subsection{Fundamental of $\SO(N)$}
\label{sec:mato}

Next we consider a set of massless scalars $\pi^a$'s transforming under the fundamental representation of $\SO (N)$ group. In CCWZ this could arise from the $\SO (N+1)/\SO (N)$ coset. The group generators satisfy the completeness relation
\bea
(T^i)_{ab } (T^i)_{cd} = \frac{1}{2}  (\delta^a_d \delta^b_c - \delta^a_c \delta^b_d)\ .\label{eq:sonc}
\eea
In this case the IR construction of the effective Lagrangian simplifies considerably due to the property
\bea
(\mt)_{ab} = \frac{1}{f^2}  (T^i)_{ac} (T^i)_{db}\ \pi^c \pi^d = \frac{1}{2f^2} \left(\<\pi| \pi \> \delta_{ab} - \pi^a\pi^b\right) \ ,
\eea
where $\mt$ is defined in Eq.~(\ref{eq:shiftall}). Denote $r \equiv \sqrt{ \< \pi | \pi \>/(2 f^2)}$, we further have
\bea
\mt^n  = r^{2(n-1)} \mt \ ,
\eea
which allows one to simplify an arbitrary function $F(\mt)$: 
\bea
[F(\mt )]_{ab} = \frac{1}{r^2} \left[ F(r^2) - F(0) \right] (\mt)_{ab} + F(0) \delta_{ab}\ .
\eea
In this case the Goldstone covariant derivative also simplifies:
\bea
|d_\mu \> &=& \frac{1}{f} F_1 (r^2) | \partial_\mu \pi \> - \frac{F_1(r^2) - 1}{2 f^3 r^2} \< \pi \partial_\mu \pi \> |\pi \>,\label{eq:sond}
\eea
and the leading two-derivative  Lagrangian becomes
\bea
\lag^{(2)} &=& \frac{1}{2} F_1^2 (r^2) \<\partial_\mu \pi| \partial^\mu \pi \> - \frac{1}{4f^2 r^2} \left[ F_1^2 (r^2) - 1 \right] \< \pi | \partial_\mu \pi \>^2.\label{eq:lnsonf}
\eea

The important observation following from Eq.~(\ref{eq:lnsonf}) is that, because of the completeness relation in Eq.~(\ref{eq:sonc}), the scalars in ${\cal L}^{(2)}$ are {\em pair-wise} contracted by the Kronecker delta. Because of the Bose symmetry, the amplitude must be symmetric in exchange of external momenta corresponding to pair-wise contracted scalars. This property agrees with that of the amplitudes soft-bootstrapped from the double-trace soft block at $\ordr (p^2)$ in Section \ref{sec:op2doub}.  Indeed we are able to match the amplitudes from Eq.~(\ref{eq:lnsonf}) with those from the double-trace soft block $S^{(2)}(1,2|3,4)$ in Eq.~(\ref{eq:p2dtsb}). For example, the 4-pt vertex given by $\lag^{(2)}$ is
\bea
-\frac{1}{12f^2} \left( \< \pi | \pi \> \<\partial_\mu \pi| \partial^\mu \pi \> -  \< \pi | \partial_\mu \pi \>^2 \right),
\eea
which generates the following vertex:
\bea
V^{(2)} (1,2|3,4) = \frac{1}{6f^2} \left[2 p_1 \cdot p_2 + 2 p_3 \cdot p_4 + (p_1 + p_2)^2\right].\label{eq:sonp24v}
\eea
Then the 4-pt soft block at $\ordr (p^2)$ is
\bea
{\cal S}^{(2)}(1,2|3,4) = \frac{1}{2f^2}  s_{12}.
\eea
Matching with Eq. (\ref{eq:p2dtsb}), we have
\bea
d_0 = \frac{1}{2}.
\eea
Similarly, using Eq. (\ref{eq:sonp24v}) and the 6-pt vertex generated by $\lag^{(2)}$:
\bea
V^{(2)} (1,2|3,4|5,6) &=&- \frac{1}{90f^2} \left[8\left( p_1 \cdot p_2 + p_3 \cdot p_4 + p_5 \cdot p_6 \right) \right.\non\\
&&\left.+ (p_1 + p_2)^2 + (p_3 + p_4)^2 + (p_5 + p_6)^2\right],
\eea
we can calculate the 6-pt amplitude which agrees perfectly with Eq. (\ref{eq:sonp26pa}).

At $\ordr (p^4)$, the number of independent operators is enumerated in Refs.~\cite{Liu:2018vel,Liu:2018qtb} for the $\SO(5)/\SO(4)$ coset, although we checked that the counting is valid for all $N$.\footnote{The non-existence of independent operator $O_4$ can be proved by using the fact that for the coset $\SO (N+1)/ \SO (N)$, all totally symmetric rank-4 tensors, with indices in the adjoint and restricted to ones associated with the ``broken generators,'' can be expressed using Kronecker deltas.} Again, using the completeness relation in Eq. (\ref{eq:sonc}) all flavor indices are contracted by Kronecker deltas: 
\bea
O_1 &=& \frac{[F_1 (r^2)]^4}{f^4} \< \partial_\mu \pi | \partial^\mu \pi \>^2 -  \frac{[F_1(r^2)]^2 \{[F_1(r^2)]^2 - 1\}}{ f^6 r^2}  \< \partial_\mu \pi | \partial^\mu \pi \> \< \pi| \partial_\nu  \pi \>^2\non\\
&& +  \frac{\{[F_1(r^2)]^2 - 1\}^2}{4 f^8 r^4} \<   \pi | \partial_\mu \pi\>^4,\label{eq:o1son}\\
O_2 &=&  \frac{[F_1 (r^2)]^4}{f^4} \< \partial_\mu \pi | \partial_\nu \pi \>^2 -  \frac{[F_1(r^2)]^2 \{[F_1(r^2)]^2 - 1\}}{ f^6 r^2}  \< \partial_\mu \pi | \partial_\nu \pi \>  \< \pi| \partial^\mu  \pi \> \< \pi| \partial^\nu  \pi \>\non\\
&& +  \frac{\{[F_1(r^2)]^2 - 1\}^2}{4 f^8 r^4} \<   \pi | \partial_\mu \pi\>^4\label{eq:o2son}.
\eea
The number of independent operators matches the number of soft blocks at $\ordr (p^4)$ in EFT$_2$, which has $d_1$ and $d_2$ as the free parameters. Furthermore, using $O_1$ and $O_2$ we calculate the 4-pt vertex to be
\bea
V^{(4)} (1,2|3,4) &=&   \frac{4}{f^2 \Lambda^2} \left[ 2C_1 p_1 \cdot p_2 \ p_3 \cdot p_4  + C_2 ( p_1 \cdot p_3 \ p_2 \cdot p_4+ p_1 \cdot p_4 \ p_2 \cdot p_3 ) \right] ,
\eea
which results in the soft block at $\ordr (p^4)$:
\bea
{\cal S}^{(4)}(1,2|3,4)&=&\frac{1}{f^2 \Lambda^2} \left[ (2 C_1 + C_2) s_{12}^2 - 2C_2 s_{13} s_{23} \right].
\eea
Comparing with Eq. (\ref{eq:sa4d}) we are able to identify
\be
 d_1 = 2C_1 + C_2\ ,\qquad d_2 = -2C_2\ .
\ee

As for the WZW term, it is shown in Section \ref{sec:wzw} that EFT$_2$ does not have a $n=5$ WZW soft block except for $N_f=5$, which corresponds to  a fundamental representation in $\SO (5)$. Indeed we show in Appendix \ref{app:allop} that there is no WZW term for the coset $\SO ( N+1)/ \SO (N)$, except for $N=5$. For the coset $\SO (6) / \SO (5)$, the WZW term can be expressed as
\bea
S_\wzw &\propto & \int d^5 y\ \vep^{\alpha \beta \gamma \delta \epsilon}\ d_\alpha^a d_\beta^b d_\gamma^c d_\delta^d d_\epsilon^e\ \vep^{abcde } .
\eea
Then the 5-pt operator in the Lagrangian is
\bea
  \frac{C_5}{f^3 \Lambda^2 } \vep^{\mu \nu \rho \sigma}  \pi^a \partial_\mu \pi^b \partial_\nu \pi^c \partial_\rho \pi^d \partial_\sigma \pi^e \vep^{abcde}.
\eea
As the flavor factor is just $\vep^{abcde}$, the partial amplitude is given by
\bea
M^{(4),a_1 a_2 a_3 a_4 a_5} (p_1, p_2, p_3, p_4, p_5) = \sum_{\sigma \in S_4} \vep^{ a_{\sigma (1)} a_{\sigma (2)} a_{\sigma (3)} a_{\sigma (4)} a_5} {\cal S}^{(4)}_- (\sigma (1), \sigma (2), \sigma (3), \sigma (4) , 5  ),\non\\
\eea
with
\bea
{\cal S}^{(4)}_- (1,2,3,4,5) = \frac{5 C_5}{f^3 \Lambda^2 } \vep (1234),
\eea
thus $c_- = 5 C_5$. For the 7-pt amplitude, there are at least 3 external states of the same flavor. Suppose $a_5 = a_6 = a_7$, the partial amplitude is defined as
\bea
M^{(4),a_1 a_2 \cdots a_7} (p_1, p_2, \cdots,p_7) = \sum_{\sigma \in S_4} \vep^{ a_{\sigma (1)} a_{\sigma (2)} a_{\sigma (3)} a_{\sigma (4)} a_5} M^{(4)}_- (\sigma (1), \sigma (2), \sigma (3), \sigma (4),\{5,6,7\}).\non\\
\eea
The 7-pt operator in the Lagrangian is
\bea
 -\frac{ 5 C_-}{21f^5 \Lambda^2} \vep^{\mu \nu \rho \sigma} \<\pi | \pi \> \pi^a \partial_\mu \pi^b \partial_\nu \pi^c \partial_\rho \pi^d \partial_\sigma \pi^e\ \vep^{abcde},
\eea
from which we can calculate the 7-pt vertex:
\bea
V (1,2,3,4,\{ 5,6,7 \} ) =- \frac{10C_-}{3f^5 \Lambda^2 } \vep (1234).
\eea
Using the 4-pt, 5-pt and 7-pt vertices, we have calculated the 7-pt partial amplitude, which exactly matches the result in Eq. (\ref{eq:so57pa}) generated by the soft recursion.

We reach the conclusion that
\be
\text{EFT$_2$ =  $\SO (N)$ Fundamental NLSM.} \non
\ee
As emphasized already, this is a new example where the partial amplitudes can be soft-bootstrapped in a simple manner. In particular, the WZW term in EFT$_2$ exists only for $N=5$, in accordance with the expectation from group-theoretic arguments.

\section{Summary and Outlook}

In this work we have considered soft bootstrapping four-derivative operators in a multi-scalar EFT which satisfies the Adler's zero condition. We systematically introduced soft blocks, the seeds of soft bootstrap, at both the leading two-derivative order and the four-derivative order. We find 7 soft blocks in total, up to $\ordr (p^4)$, which are summarized in Table \ref{tab:softblocks}. A consistent EFT can be bootstrapped starting from either $c_0$ or $d_0$ at $\ordr (p^2)$, but not both. Going up to $\ordr (p^4)$, two EFT's can be constructed using the relevant soft blocks:
\begin{align}
{\rm EFT}_1 &= \{ c_0, c_1, c_2, d_1,d_2, c_- (N_f\ge 5) \} \ , \nonumber \\
 {\rm EFT}_{2} &= \{d_0, d_1, d_2, c_-(N_f=5) \} \ , \nonumber
\end{align}
where we have indicated the $N_f$ dependence of the WZW term, which arises from the Bose symmetry requiring the amplitudes to be invariant under exchange of momentum labels corresponding to identical bosons.

\begin{table}[t]
\begin{center}
\begin{tabular}{|c|c|c|c|}
\hline
    \multicolumn{2}{|c|}{ }                   &   Single-trace  & Double-trace \\
\hline \hline                      
\multicolumn{2}{|c|}{$\ {\ordr}(p^2)\ $ }  & $\ {\cal S}^{(2)}(1,2,3,4) = c_0\ s_{13}/f^2\ $ & $\ {\cal S}^{(2)}(1,2|3,4) = d_0\ s_{12}/f^2\ $\\
\hline
\multirow{3}{*}{$\ \ordr (p^4)\ $} &\multirow{2}{*}{$\ $ P-even$\ $} & $\ {\cal S}_1^{(4)}(1,2,3,4) = c_1\ s_{13}^2/(\Lambda^2 f^2)\ $ &  $\ {\cal S}_1^{(4)}(1,2|3,4) = d_1\ s_{12}^2/(\Lambda^2 f^2)\ $ \\
\cline{3-4}
 & & $\ {\cal S}_2^{(4)}(1,2,3,4) = c_2\ s_{12}s_{23}/(\Lambda^2 f^2)\ $ &  $\ {\cal S}_2^{(4)}(1,2|3,4) = d_2\ s_{13}s_{23}/(\Lambda^2 f^2)\ $ \\
\cline{2-4}
& $\ $ P-odd$\ $ & $\ {\cal S}_-^{(4)}(1,2,3,4,5) = c_-\ \vep (1234) /(\Lambda^2 f^3)\ $ & $-$ \\
\hline
\end{tabular}
\caption{\em Soft Blocks up to $\ordr (p^4)$. Each soft block comes with an unknown free parameter. At $\ordr (p^2)$ one must choose between $c_0$ or $d_0$, which can be absorbed into the normalization of $f$. We have also indicated the parity of the $\ordr(p^4)$ soft block.\label{tab:softblocks}}
\end{center}
\end{table}

We explicitly matched these two EFT's to the coset construction of NLSM effective Lagrangians:
\begin{align}
\text{EFT}_1 &= \SU(N)\times \SU(N)/\SU(N) \ , \nonumber \\
\text{EFT}_2 &= \SO(N+1)/\SO(N) \nonumber \ .
\end{align}
Therefore, massless scalars in EFT$_1$ transform as the adjoint of the unbroken $\SU (N)$, while in EFT$_2$ they transform as the fundamental representation of $\SO (N)$. At the leading two-derivative order, both $\SU (N)$ adjoint and $\SO (N)$ fundamental have a single operator that is nonlinear in $1/f$. The overall coefficient of the nonlinear operator is fixed by requiring a canonically normalized kinetic term for the massless scalars. When expanding in $1/f$, the nonlinear operator gives rise to vertices that  carry two derivatives and an increasing number of scalar fields. Coefficients in front of these vertices, at each order in $1/f$, are completely determined by soft bootstrap, using the two-derivative soft blocks. At $\ordr (p^4)$, there are multiple operators nonlinear in $1/f$ and each carrying its own Wilson coefficient, which is incalculable in the IR. In soft bootstrap the Wilson coefficient arises from the free parameter associated with each soft block at $\ordr (p^4)$, and there is a one-to-one correspondence between the soft blocks and the four-derivative nonlinear operators. Again when expanding in $1/f$, vertices in each nonlinear operator are fixed by soft bootstrap.

For the WZW term in the coset, its existence relies on the anti-symmetric rank-5 tensor in the coset involved. Group-theoretic consideration suggests there is no WZW term in the $\SU (N)$ adjoint theory for $N = 2$ and in the $\SO (N)$ fundamental theory for $N\neq 5$. Remarkably, soft bootstrap is able to reproduce these results by considering the number of flavors involved and a novel application of Bose symmetry.

Our success of extending the soft bootstrap program to $\ordr (p^4)$ of NLSM strongly suggests that, by using the soft recursion relations, we should be able to construct the full EFT to all orders in the derivative expansion, at least for certain cosets. An advantage of such a method is that, at a given order in derivative expansion, it is remarkably easy to find the general set of independent operators: all we need to do is to enumerate all soft blocks that satisfy certain ordering properties as well as the Adler's zero condition, and make sure that they lead to consistent higher-pt amplitudes. Therefore, we are able to avoid applying the relations of nonlinear symmetries and  equations of motion to reduce the number of independent operators in the Lagrangian, which become increasingly complicated when we go to higher orders. In this sense, our work is similar in spirit to recent attempts of classifying higher dimensional operators in the standard model and beyond the standard model EFTs using an amplitude basis \cite{Shadmi:2018xan,Ma:2019gtx}. There also exist algorithms that enumerate independent operators in NLSM by utilizing the Hilbert series \cite{Henning:2017fpj}, and one should explore how they are related to the soft blocks.

Going to even higher orders in derivative expansion, one can see that new complications may appear in soft bootstrap. One example is the $\U (1)$ decoupling relation, which is evident for the $\ordr (p^2)$ amplitudes. When soft-bootstrapping amplitudes at $\ordr (p^4)$, we only need the decoupling relation at $\ordr(p^2)$ because in a factorization channel at least one of the two sub-partial amplitudes is at $\ordr (p^2)$. At $\ordr (p^6)$, nevertheless, one or both of the two sub-amplitudes  can  be at $\ordr (p^4)$ and one would need to prove first that $\U (1)$ decoupling relations still holds at $\ordr (p^4)$. It remains to be seen if this is the case. Another complication arises from the fact that, in general, the 6-pt amplitudes at $\ordr (p^6)$ are not soft constructible anymore by simple power counting, as the integrand in Eq.~(\ref{eq:cauc}) does not vanish at $z=\infty$. There is a very good reason for this behavior, however, as we can construct 6-pt soft blocks at $\ordr (p^6)$ that satisfy the Adler's zero condition; thus they need to be given as the input for soft bootstrap and cannot be constrained  using the soft recursion. Moreover, soft blocks at higher orders in the derivative expansion have enhanced soft limits. This might allow us to increase the power of $z$ in the soft factor $F_n (z)$ given by Eq.~(\ref{eq:sffn}), so as to make the integrand vanishes at $z=\infty$ in soft recursion relation. It will be interesting to see how the details work out at $\ordr (p^6)$.

There are many more future directions to consider. In particular, quantum field theories with matter content in the adjoint of $\SU (N)$ has been studied heavily in the scattering amplitudes community, because the color/flavor factor factorizes simply in each of the kinematic factorization channel, resulting in simple relations between the full and the partial amplitudes. The fundamental of $\SO(N)$ is a new example of quantum field theories enjoying such a nice property as in the $\SU(N)$ adjoint theory. It is possible that they may be related, e.g. by dimensional reduction \cite{Chen:2014cuc}. At $\ordr (p^2)$, the flavor factor in $\SU (N)$ adjoint theory can be written as a single trace over group generators, resulting in two special properties
\begin{itemize}

\item The partial amplitude is invariant under cyclic permutation of external particles.

\item The factorization channel can only arise from adjacent momenta.

\end{itemize}
Neither property holds at $\ordr (p^4)$ in $\SU (N)$ adjoint theory, because of operators containing double trace. For the fundamental of $\SO (N)$, both properties fail already at $\ordr (p^2)$, let alone $\ordr (p^4)$. It would be interesting to study whether the double-copy structure carries over to $\ordr (p^4)$ in $\SU (N)$ adjoint theory and/or to the $\SO(N)$ fundamental theory at all. Ref. \cite{Elvang:2018dco} pointed out that there are no single trace 4-pt soft blocks that satisfy the fundamental BCJ relation, thus if there is some kind of double-copy structure at $\ordr (p^4)$ in the $\SU (N)$ adjoint NLSM, it is not in a form that we naively expect it to be. A related question is whether there exists the CHY representation, which makes the double copy structure manifest at $\ordr (p^2)$ for the $\SU(N)$ adjoint, for $\ordr (p^4)$ operators and for the $\SO (N)$ fundamental theory.

Another interesting direction is related to the recent proposal to directly interpret tree-level amplitudes as canonical forms associated with the positive geometry in the space of kinematic invariants \cite{Arkani-Hamed:2017mur}. Geometric interpretations are given for a variety of theories, including pions transforming in the adjoint of $\SU (N)$ theory at the leading two-derivative order. It remains to be seen whether the $\ordr (p^4)$ amplitudes in $\SU (N)$ and/or the amplitudes in $\SO (N)$ theories can be incorporated in such a narrative. In particular, the scattering form proposed so far is projective. It was remarked earlier that the general solutions to $a_i$'s, defined in the all-line shift in soft recursion relation, are also defined projectively (and enjoy a shift symmetry.) It is natural to wonder if these shift parameters can be given a meaning in the projective geometry in the space of kinematic invariants.

Last but not least, it is intriguing that a purely IR approach like the soft bootstrap could make statements on the existence of the WZW term, or the lack thereof, which relies on group-theoretic arguments previously. One could further ask whether it is possible to {\em derive} properties of the Lie group involved in the EFT's based simply on the notion of discrete ordering in soft bootstrap. For example, would it be possible to derive in EFT$_1$ that the number of flavors could only be $N^2-1$? If we only assume cyclic properties of the partial amplitudes, can one deduce that the flavor factors must satisfy a ``Jacobi identity," thereby establishing its group nature?

We leave the study of these questions for future investigations.

\begin{acknowledgments}

We thank John Joseph Carrasco and Laurentiu Rodina for useful discussions and comments on the manuscript, and Yu-tin Huang for thoughtful comments. Z. Y. would like to thank Marios Hadjiantonis, Callum R.T. Jones, Shruti Paranjape, Andreas Trauter and Jaroslav Trnka for helpful discussions. This work is supported in part by the U.S. Department of Energy under contracts No. DE-AC02-06CH11357 and No. DE-SC0010143.

\end{acknowledgments}

\begin{appendix}

\section{Multi-trace Flavor-ordered Partial Amplitudes}
\label{app:multi}

Refs.~\cite{Cheung:2015ota,Cheung:2016drk} only considered partial amplitudes of a single flavor factor as shown in Eq. (\ref{eq:stfod}); namely, the flavor factor is a single trace of generators $T^a$. However, in general the flavor factor can be a product of $t$ traces, thus we can define the corresponding flavor-ordered amplitudes using
\bea
M^{a_1 \cdots a_{n} } (p_1, \cdots, p_{n})& \equiv&
\sum_{t=1}^{\lfloor n/2 \rfloor } \sum_{ l}  \sum_{\sigma \in S_{n} / S_{n;l} } \left( \prod_{i=1}^{t} \mathcal{C}^{a_{\sigma  (l_{i-1} +1)} \cdots  a_{\sigma (l_i)}}   \right)  M_{\sigma ; l} (p_1, \cdots , p_n),\label{eq:mttfo}
\eea
where $l = \{l_0, \cdots, l_t \}$ labels the possible partition of ordered indices $\{ 1,2, \cdots ,n\}$ into $t$ subsets, with the requirement of $l_0 = 0$, $l_t = n$ and $l_{i+1} - l_i \le l_{i+2} - l_{i+1}$, $i=0,1, \cdots ,t-2$; $S_{n; l}$ are permutations of the indices $\{ 1,2, \cdots ,n\}$ that leave the flavor factor invariant. Similar to the case of single-trace amplitudes, we will denote the multi-trace flavor-ordered amplitude $M_{\sigma ; l} (p_1, \cdots , p_n)$ as
\bea
M ( \sigma (1), \cdots, \sigma (l_1) | \sigma(l_1 + 1), \cdots , \sigma (l_2 ) | \cdots | \sigma (l_{t-1} + 1), \cdots \sigma (n) ).
\eea
For the amplitude
\bea
M (1,2, \cdots l_1|l_1 + 1, \cdots , l_2 | \cdots | l_{t-1} + 1, \cdots, n),
\eea
it is invariant when we  do the cyclic permutation separately for the sets of indices $\{1,2, \cdots, l_1\}$, $\{ l_1+1, \cdots, l_2 \}$ and so on. Furthermore, if $l_{i+1} - l_i = l_{i+2} - l_{i+1}$, exchanging the sets $\{ l_i+1, \cdots ,l_{i+1} \}$ and $\{ l_{i+1} + 1, \cdots , l_{i+2} \}$ will also leave the amplitude invariant.

\section{The IR Construction of Effective Lagrangians at ${\cal O}(p^4)$}
\label{app:allop}

In this appendix we derive  the effective Lagrangians of NLSM using the IR construction established in Refs.~\cite{Low:2014nga,Low:2014oga}. The leading two-derivative operator was considered in Section \ref{app:infc}, where we have set the notation and the basis of generators for the unbroken group $H$. The building blocks, $d^a_\mu$ and $E_\mu^i$, are given in Eqs. (\ref{eq:irdefd}) and (\ref{eq:irdefe}). We assume the number of massless scalars is $n$ and the number of generators in $H$ is $N_g$,  so that the range of indices are
\be
\{a, b, \cdots\} = \{1, \cdots, n\}\ , \qquad \{i, j, \cdots\}=\{1,\cdots, N_g\}\ ,
\ee
where $\{a, b,\cdots\}$ run in the linear representation $R$ furnished by the massless scalars $|\pi\rangle$ and $\{i,j,\cdots\}$ are indices in the adjoint of $H$. We have chosen a basis so that generators in the representation $R$ is anit-symmetric and purely imaginary,
\be
(T^i)_{ab} = - (T^i)_{ba}\ , \qquad (T^i)_{ab}^* = - (T^i)_{ab} \ .
\ee
Recall that the generators $T^i$ satisfy the Lie algebra
\be
[T^i, T^j]= i f^{ijk} T^k\ ,
\ee
where $f^{ijk}$ is the structure constant. The {infrared} data available to us in the low energies are therefore $f^{ijk}$ and $(T^i)_{ab}$.

We will define two sets of $(N_g+n)\times(N_g+n)$ Hermitian matrices $\mathsf{X}$ and $\mathsf{T}$ 
\bea
\mathsf{X}^a&=& \left[
\begin{array}{c|c}
{\Large \varnothing}  &\  A^a\  \,  \\
\hline
(A^a)^\dagger & {\Large \varnothing} 
\end{array} \right] \ , \quad  a=1,\cdots, n \ , \\
\mathsf{T}^i&=&
 \left[
\begin{array}{c|c}
\ B^i\  & {\Large \varnothing}   \,  \\
\hline
 {\Large \varnothing} & \ C^i\  
\end{array} \right]  \ , \quad i=1, \cdots, N_g \ ,
\eea
where $A^a$ is an $N_g\times n$ matrix, $B^i$ an $N_g\times N_g$ matrix and $C^i$ an $n\times n$ matrix:
\be
(A^a)_{ib}= -(T^i)_{ab} \ , \qquad (B^i)_{jk} = -i f^{ijk} \ , \qquad (C^i)_{ab} = (T^i)_{ab} \ .
\ee
These matrices are defined entirely using IR data. However, it is possible to make connection with the coset construction by the identification
\be
(T^i)_{ab} = - i f^{iab} \ ,
\ee
using which one sees $\mathsf{X}^a$ and $\mathsf{T}^i$ are nothing but the ``broken'' and ``unbroken'' generators in the CCWZ construction.

Armed with the IR definition of $\mathsf{X}^a$ and $\mathsf{T}^i$, one can now proceed to define the Cartan-Maurer one-form in the IR,
\be
\mathsf{\Omega}\equiv e^{i\pi^a \mathsf{X}^a/f} \ , \quad \qquad \mathsf{\Omega}^\dagger \partial_\mu \mathsf{\Omega} = i \left(d_\mu^a\, \mathsf{X}^a + E_\mu^i\, \mathsf{T}^i \right)\ .
\ee
Under the nonlinear shift symmetry in Eq.~(\ref{eq:shiftall}), they transform covariantly and inhomogeneously as shown in Eqs.~(\ref{eq:dmutransf}) and (\ref{eq:eitransf}). Using the automorphism $\mathsf{X}^a \to -\mathsf{X}^a$ and $\mathsf{T}^i \to \mathsf{T}^i$, we have 
\bea
d_\mu &=& -\frac{i}{2} [\mathsf{\Omega}^\dagger \partial_\mu \mathsf{\Omega} -\mathsf{\Omega} \partial_\mu \mathsf{\Omega}^\dagger ]\ ,\label{eq:gdefd}\\
E_\mu &=&-\frac{i}{2} [\mathsf{\Omega}^\dagger \partial_\mu \mathsf{\Omega} + \mathsf{\Omega} \partial_\mu \mathsf{\Omega}^\dagger ]\ .\label{eq:gdefe}
\eea
Using these expressions we can work out the form of $d^a_\mu$ and $E^i_\mu$ explicitly, by calculating the derivative of the exponential map:
\bea
e^{-X (x)} \partial_\mu e^{X(x) } = \frac{1- e^{-\Adj_X}}{\Adj_X } \partial_\mu X(x) ,
\eea
where $\Adj_X Y \equiv [X, Y]$, and
\bea
\frac{1- e^{-\Adj_X}}{\Adj_X } = \sum_{k=0}^\infty \frac{(-1)^k }{(k+1)! } (\Adj_X)^k.
\eea
Combining with Eqs. (\ref{eq:gdefd}) and (\ref{eq:gdefe}), we arrive at the expressions for $d_\mu^a$ and $E^i_\mu$ in Eqs. (\ref{eq:irdefd}) and (\ref{eq:irdefe}). Two important identities follow from Eqs. (\ref{eq:gdefd}) and (\ref{eq:gdefe}),
\bea
\nabla_{[\mu} d_{\nu] } &=& 0\ ,\label{eq:covd0} \\
E_{\mu \nu} &\equiv& -i [ \nabla_\mu, \nabla_\nu]=  -i[d_{\mu}, d_{\nu}]\ ,
\eea
where $\nabla_\mu d_\nu \equiv \partial_\mu d_\nu + i [E_\mu, d_\nu]$.
In the geometric construction of a symmetric coset  the identities follow from the Maurer-Cartan equation \cite{DHoker:1995mfi}.

The leading two-derivative Lagrangian is already presented in Eq.~(\ref{eq:nlsmir}). Using $d_\mu$, $\nabla_\mu d_\nu$ and $E_\mu$  we write down the following 8 parity-even, $\ordr (p^4)$ operators,
\bea
O_1 &=& \left[ {\rm tr} (d_\mu d^\mu)\right]^2,\label{eq:o4nt1}\\
O_2 &=& \left[ {\rm tr} (d_\mu d_\nu)\right]^2,\label{eq:o4nt2}\\
O_3 &=& \tr ( [ d_\mu, d_\nu]^2 ) = - \tr ( E_{\mu \nu}^2 ),\\
O_4 &=& \tr ( \{ d_\mu, d_\nu \}^2 ),\\
O_5 &=& \tr ( d_\mu d^\mu \nabla_\nu d^\nu ),\\
O_{6} &=& \tr ( d_\mu \nabla^\mu \nabla^\nu d_\nu )\ , \\
O_7 &=& \tr (  d_\mu d_\nu  \nabla^\mu d^\nu ) = - \tr (  d_\mu d_\nu  \nabla^\nu d^\mu )\ ,\\
O_8 &=& \tr ( d_\mu \nabla_\nu \nabla^{\mu} d^\nu )= - \tr ( d_\mu \nabla_\nu \nabla^{\nu} d^\mu )\ .
\eea
Using integration-by-parts, one can show that $O_7$ is not independent of $O_5$, up to a total derivative, and that $O_8$ is a linear combination of $O_3$ and $O_6$. We can choose to eliminate $O_7$ and $O_8$ from the list. Furthermore, the equation of motion from the leading two-derivative operator is
\be
\nabla_\mu d^\mu = 0 \ ,
\ee
which implies $O_{5}$ and $O_6$ vanish up to ${\cal O}(\partial^4)$. So in the end we are left with four parity-even operators, $O_i, i=1,\cdots, 4$, in general. However, as emphasized in Section \ref{sec:mat}, the number of independent operators could be further reduced, depending on the specific group structure, such as in $\SU (2), \SU(3)$ and $\SO(N)$.

The operators considered so far are those that are invariant under the shift symmetry. There is an operator that varies by a total derivative in the Lagrangian, which is the Wess-Zumino-Witten term \cite{Wess:1971yu,Witten:1983tw}. To write down the Lagrangian density for the WZW term requires compactifying the spacetime to a 4-sphere $M_4$ and extending $\pi^a(x) \to \tilde{\pi}^a(x, s)$ such that $\tilde{\pi}^a(x,1)=\pi^a(x)$ and $\tilde{\pi}^a(x,0)=0$. One then defines a 5-ball $B_5$ with boundary $M_4$ and coordinates $y^\alpha=\{x^\mu, s\}$. The WZW action can be written as
\be
\label{eq:actionwzw}
S_{\wzw} \propto \int d^5y\ \vep^{\alpha\beta\gamma\delta\sigma}\ \omega^{abcde}\ d_\alpha^a d_\beta^b d_\gamma^c d_\delta^d d_\sigma^e \ ,
\ee 
where $d_\alpha^a$ is the Goldstone covariant derivative in Eq.~(\ref{eq:irdefd}), suitably extended to $B_5$. The totally anti-symmetric Levi-Civita tensor forces the rank-5 invariant tensor  $\omega^{abcde}$ to be totally anti-symmetric as well. Group-theoretically, the existence of WZW action now is related to the existence of a rank-5 totally anti-symmetric tensor in the particular representation $R$ of $H$ that is furnished by $\pi^a$, which is given by the fifth de Rham cohomology group $H^5$ \cite{DHoker:1994rdl,DHoker:1995mfi}.

The fifth de Rham cohomology group of the symmetric space of simple Lie groups is well-known. For $\pi^a$'s furnishing the adjoint representation of $\SU (N)$ group, they can be thought of as coordinates parameterizing the coset space $\SU(N)\times\SU(N)/\SU(N)$. For $N\ge 3$, $H^5(\SU(N),R)$ has a single generator which is precisely the integrand in Eq.~(\ref{eq:actionwzw}) \cite{DHoker:1994rdl,DHoker:1995mfi}. The other case of interest is when $\pi^a$'s furnish the fundamental representation of $\SO(N)$. In this case $\pi^a$'s parameterize the coset $\SO(N+1)/\SO(N)$. It turns out that $H^5(\SO(N),R)$ is zero except for $N=6$, which can be understood from the local isomorphism $\SO(6)\approx \SU(4)$ \cite{DHoker:1994rdl,DHoker:1995mfi}. We conclude that
\bea
\text{Adjoint\ of\ $\SU(N)$}&:& \ S_{\wzw} \neq 0 \qquad \text{for} \qquad N \ge 3 \ ,\nonumber\\
\text{Fundamental\ of\ $\SO(N)$}&:& \ S_{\wzw} \neq 0 \qquad \text{for} \qquad N=5 \ .
\eea

If we expand Eq.~(\ref{eq:actionwzw}) in $1/f$, it contains a series of local operators on the spacetime:
\be
S_{\wzw} \propto \int d^4x \ \vep^{\mu\nu\rho\sigma}\  \omega^{abcde}\ \pi^a\partial_\mu\pi^b \partial_\nu\pi^c \partial_\rho\pi^d\partial_\sigma\pi^e + \cdots \ ,
\ee
which contribute to $(5+2n)$-pt amplitudes.

\end{appendix}

\bibliography{references_amp}

\end{document}